\documentclass[12pt]{article}
\pdfoutput=1

\usepackage{hyperref}
\hypersetup{colorlinks,linkcolor=blue,citecolor=blue,bookmarks,bookmarksnumbered}

\usepackage[utf8x]{inputenc}
\usepackage[scaled=0.85]{helvet}
\usepackage{amsmath,amssymb,accents,mathrsfs,XoohmE}
\usepackage{graphicx,color}
\usepackage{placeins}
\usepackage{amsmath}
\usepackage{array}
\usepackage{nicematrix}
\usepackage{makecell}
\usepackage{algorithm}
\usepackage[noend]{algpseudocode}
\usepackage[nosort,nomove,compress]{cite}

\newlength\mylen
\graphicspath{{Figures/}}

\newcommand{\bm}[1]{\mbox{\boldmath$#1$}}

\definecolor{Hey}{rgb}{.9,.05,.4}
\definecolor{orange}{rgb}{1,.5,0}
\definecolor{plum}{rgb}{.4,0,.6}
\definecolor{R}{rgb}{1,0,0}
\definecolor{G}{rgb}{0,1,0}
\definecolor{B}{rgb}{0,0,1}

\newcolumntype{M}[1]{>{\centering\arraybackslash}m{#1}}
\numberwithin{equation}{section}

\begin{document}

\thispagestyle{empty}
\noindent{\small
\hfill{$~~$}  \\ 
{}
}
\begin{center}
{\large \bf
Adinkras \& Genomics in Sixteen Color Systems (I) \\
\vskip4pt
$~~$  
}  \\   [8mm]
{\large {
Jeth Arunseangroj\footnote{jarunseangro@student.ethz.ch}${}^{, a}$,
Jude Bedessem\footnote{judejb2@illinois.edu}${}^{, b}$,
S.\ James Gates, Jr.\footnote{gatess@umd.edu}${}^{, c, d}$, \\
Gabriel Yerger\footnote{gabeyerger@gmail.com}${}^{, e}$
}}

\emph{
\centering
$^{a}$ETH Z\"{u}rich, Department of Physics
\\[1pt]
8093 Z\"{u}rich, Switzerland\\[10pt]
$^{b}$University of Illinois, Department of Physics\\[1pt]
Loomis Laboratory of Physics, Urbana, IL 61801, USA\\[10pt]
and
\\[10pt]
$^{c}$University of Maryland, Department of Physics
\\[1pt]
J.\ S.\ Toll Hall, Room 1117,
College Park, MD 20742-4111, USA
\\[10pt]
$^{d}$University of Maryland, School of Public Policy
\\[1pt]
T. Marshall Hall, Room 2223,
College Park, MD 20742-5035, USA\\[10pt]
$^{e}$Brown University, Department of Physics
\\[1pt]
Barus \& Holley, Room 545, Providence, RI 02912}
 \\\vfill
{ ABSTRACT}\\[05mm]
\parbox{142mm}{\parindent=2pc\indent\baselineskip=14pt plus1pt
Motivated by the search for embedded on-shell supermultiplets in higher dimensional off-shell theories, we investigate several 16-color supermultiplets and their topology. An Adinkra's topology is known to be equivalent to $(\mathbb{Z}_2)$-quotients of an N-cube. This is revisited with the focus of closing the off-shell problem for the 4D $\mathcal{N} = 4$ Maxwell supermultiplet.
} \end{center}
\vfill
\vfill
\clearpage

\newpage
{\hypersetup{linkcolor=black}
\tableofcontents
}

\newpage

\section{Introduction}
\label{sec:Intro}

$~~~~$ In this work, we investigate the proposition that the off-shell problem for the 4D, N=4 Maxwell supermultiplet
offers a direction for probing made tractable with non-conventional new mathematical and computational tools. 
To restate the off-shell problem, we look for a set of auxiliary fields which, when introduced to augment the on-shell theory in question, produces a theory which closes off-shell. We emphasize that this can
conceptually be done, {\it {without making an any priori assumptions about an action nor any symmetry group representation provided by the fields.}}

The conventional ruling out of such a possibility has been the accepted wisdom about this by means of a counting argument \cite{siegel_off-shell_1981} termed the ``no-go theorem," however, we have continued to investigate this question on the premise that if a key assumption is relaxed, the accepted conclusion might be altered.

However, there is a simple counter argument to the
strong interpretation of the ``no-go theorem."  If
one begins with a superspace Yang-Mills superconnection, then clearly all of the component
fields are contained within it. One can impose a constraint that
defines the vectorial superconnection in terms of the
spinorial one and this leaves the spinorial connection
as the primary object to study.  If {\it {no further constraints are placed upon the spinorial superconnection}} then two conditions must hold:
\newline \indent
$\cdot$ all of the component fields of the on-shell
theory are contained in this superfield, and
\newline \indent
$\cdot$ the supersymmetry algebra must close without
the use of any equations of motion.

Determining the detailed component field content of this structure is one part of the version of the ``off-shell 4D, $\cal N$ =4
Maxwell Theory'' that is the final goal of our work.
The second part is to determine if there is a smaller
subset of component fields within this structure that
also satisfies the two conditions stated above.

Subsequently, the existence of an off-shell realization of the 4D, $\mathcal{N}=4$ Maxwell theory was motivated and conjectured \cite{calkins_think_2015,gates_extended_2015}, and an algorithm for deriving such a representation through its Adinkra network description was outlined. While this algorithm serves well for lower-dimensional cases, the compute time of a linear solver for n independent variables is $\mathcal{O}(n^3)$, so the case of N=16, d=128, with $\text{N}\times\text{d}$ independent variables is nearing intractability. In this work, we show that a separate route can be traveled to obtain a ruling on the possibility question, and inspire a more viable derivation of auxiliary fields.

We have long been compelled to approach the off-shell problem through innovative mathematics and the search for symmetries otherwise undetectable. We hence study and explore representation theories more fitting for the cultivation of such mathematical tools, as was the purpose of the GAAC (Garden Algebra/Adinkra/Codes) Program. The closure of the supersymmetry algebra without equations of motion is well-known to be indicated by a closure of the ``$\mathcal{GR}(d, N)$ algebra" or ``Garden algebra": the general real algebra of extension N and dimension d, a real Clifford algebra defined on the transformation rules of the compactified 0-Brane theory. The algebra is thus defined on the L and R matrices of the corresponding Adinkra:
\begin{align}
    \begin{split}
        \textbf{L}_\text{I}\textbf{R}_\text{J} + \textbf{L}_\text{J}\textbf{R}_\text{I} \;=\; 2\delta_{\text{{IJ}}}\textbf{I}\;;\\
        \textbf{R}_\text{I}\textbf{L}_\text{J} + \textbf{R}_\text{J}\textbf{L}_\text{I} \;=\; 2\delta_{\text{{IJ}}}\textbf{I} \;;\label{eqn:grdn}
    \end{split}
\end{align} 
\noindent
whereby constructing block-matrices, we recover the presentable Clifford algebra form:
\begin{gather}
    \boldsymbol{\Gamma}_\text{I} \;=\;\begin{bmatrix}
    	0 & \textbf{L}_\text{I}\\
    	\textbf{R}_{\text{I}} & 0
    \end{bmatrix}\;; \\
    \vspace{3pt}\boldsymbol{\Gamma}_{\text{I}}\boldsymbol{\Gamma}_{\text{J}} + \boldsymbol{\Gamma}_{\text{J}}\boldsymbol{\Gamma}_{\text{I}} = 2\delta_{\text{{IJ}}}\textbf{I}\;;\label{eqn:grdn2}
\end{gather}
\noindent
where $\boldsymbol{\Gamma}_\text{I}$ can be now thought as the colored and signed adjacency matrices of an Adinkra. These representation tools were born naturally in context to the 1D N-extended Spinning Particle multiplet in \cite{crombrugghe_supersymmetric_1983,gates_theory_1995,gates_theory_1996}, and were explicitly described in context to representation theory and the off-shell problem in \cite{gates_jr_fundamental_2001,gates_jr_when_2003}. Adinkras were then introduced as a means to make further strides in a complete representation theory, and in answering the long-standing auxiliary field problem for 1D SUSY \cite{faux_adinkras_2005}.

Adinkras, of which we will see several examples in this work, are graphical symbols which represent the field components of a theory and their behavior under the action of the supersymmetry transformation operator(s). From the time of their invention, they have led the way in building a comprehensive representation theory for classes of supermultiplets, and have formed a central part of what has been titled in a number of works--SUSY ``Genomics" \cite{gates_jr_4d_2009,gates_4d_2012,chappell_4d_2013,mak_4d_2019}. 

Adinkras, their matrix representations, and Garden algebras have enabled and inspired many new mathematical and computational approaches in SUSY representation theory; including but not limited to: the discovery of the holoraumy and its implications for SUSY Holography \cite{gates_clifford-algebraic_2015,gates_infinite-dimensional_2022}, the classification of supermultiplets by their topology and chromotopology using binary codes \cite{doran_relating_2008,doran_topology_2008,doran_codes_2011}, the discovery of HYMNs (Height Yielding Matrix Numbers) to detect isomorphisms between Adinkras and the predict the closure of the Garden algebra \cite{burghardt_adinkra_2012,douglas_automorphism_2015,bristow_note_2022}, investigations on the role of polytopes--namely the permutahedron--in off-shell SUSY \cite{cianciara_300_2021}, and in the properties of graph matrices and spectra of Adinkras \cite{iga_eigenvalues_2023}. A deeper look at large N-extended SUSY warrants the use of Adinkras because they provide a foundation by which all SUSY theories can be composed, and thus serve as the abacus for SUSY combinatorics. This work, being in likeness of prior ``Genomics" works, draws from the many approaches which were inspired by Adinkras, and strides further towards a complete understanding of 16-color systems. Both Adinkra technology, like codes and graph theoretical methods, as well as polytopic approaches are used to explore the ``genome" of 1D, $N=16$ supermultiplets. 

In Section \ref{sec:hoppers}, we give a detailed introduction to Hopping operators through the lens of recent concepts: the arrangement of off-shell supermultiplets on a permutahedron, and the partition of such supermultiplets into equivalence classes called ``Diadems". 

We proceed in Section \ref{sec:topologies} with a review of binary doubly-even codes and their role in classifying Adinkras. With an enhanced understanding of Adinkras through the lens of codes, we report on the topology of the 1D, N=16 Spinning Particle Multiplet. We then build on the previous literature on codes through a discussion enabled by Hopping operators.

In Section \ref{sec:maxwell}, we present several on-shell supermultiplets. These include: the 4D, N=4 Maxwell supermultiplet, 10D, $\mathcal{N}=1$ Maxwell supermultiplet, the 4D, $\mathcal{N}=4$ Maxwell supermultiplet with $\mathcal{N}=1$ degree off-shell closure, and the 4D, $\mathcal{N}=4$ Vector-Tensor supermultiplet with $\mathcal{N}=2$ degrees off-shell closure. We describe newly developed computational methods for deriving the L and R-matrix representations of these theories as well as their Hopping operators, which are included in the Appendix. 

In Section \ref{sec:embedding}, we enter the off-shell conversation, and state that by relating two theories by a quotient, they are thereby ruled topologically incompatible and do not admit any solutions to the off-shell problem. This conclusion is reached for the latter two theories: the 4D, $\mathcal{N}=4$ Maxwell supermultiplet and 4D $\mathcal{N}=4$ Vector-Tensor supermultiplet, by comparison to the Spinning Particle theory. We note that a similar conclusion evades the former two theories which are completely on-shell and have no additional auxiliary fields. In light of this, an alternative route of searching for a solution is proposed.

Finally, we conclude with remarks surrounding the present and future research on this front.

\section{Hopping Operators, Abnormal Cosets, and the Role of Polytopes}
\label{sec:hoppers}

$~~~~$ Previously, the set of all 1D, $N=4$ supermultiplets have been studied extensively \cite{cianciara_300_2021}, with particular interest placed on the permutahedron and ``magic numbers"--the sum of the shortest distances between one vertex with all others in the subset. The set of all 8-color supermultiplets have been studied just as well, with the introduction of abnormal cosets, ``hoppers", and a detailed decomposition of the $\mathbb{S}_8$ permutahedron \cite{cianciara_cal_2023}. For introductory purposes, we first illustrate how each of these objects arise in the structure of $\mathcal{GR}(4,4)$.

Begin by considering the set of all 4-color multiplets, and their permutation matrices labeled on the S$_4$ permutahedron. This labeling can be found by calculating the L and R matrices of each theory, and decomposing these matrices into sign factors $\bm{\mathcal{S}}_{\hspace{1pt}\widehat{\text{I}}}$ and permutations $\bm{\mathcal{P}}_{\hspace{1pt}\widehat{\text{I}}}$ using the following convention:
\begin{gather}
	\textbf{L}_{\hspace{1pt}\widehat{\text{I}}} = \bm{\mathcal{S}}_{\hspace{1pt}\widehat{\text{I}}}\bm{\mathcal{P}}_{\hspace{1pt}\widehat{\text{I}}}\;.\label{eqn:Lconvention}
\end{gather}

Each multiplet consists of 4 nodes which partition the permutahedron into 6 slices. The partitioned set which represents these slices was termed ``Kevin's Pizza", and is denoted by $\text{KP}_i{}^{(4)}$. As mentioned, each slice ``i" of $\text{KP}_i{}^{(4)}$ represents a different supermultiplet's permutation matrices; the correspondence of these supermultiplets with nodes on the permutahedron are shown below, as well as the $\text{KP}_i{}^{(4)}$ diagram.

\begin{figure}[ht!]
	\centering
	\includegraphics[width = 0.4\textwidth]{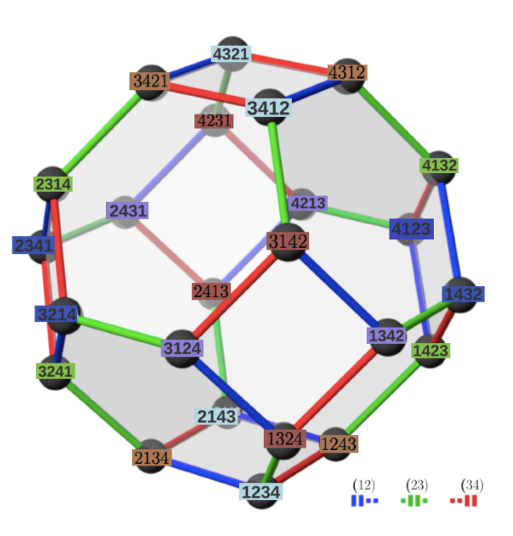}
	\hspace{-0.25cm}\raisebox{0.9cm}{\includegraphics[width = 0.35\textwidth]{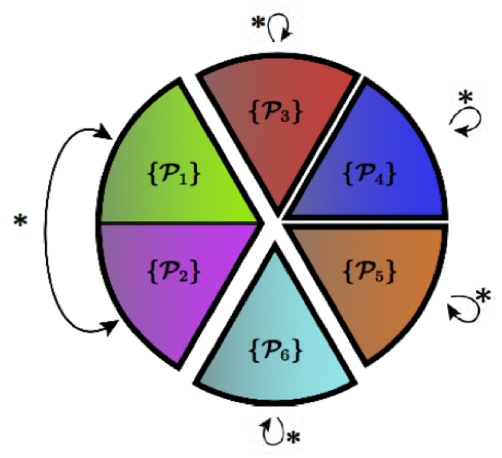}}\hspace{0.25cm}
	\raisebox{0.9cm}{\includegraphics[height = 0.3\textwidth]{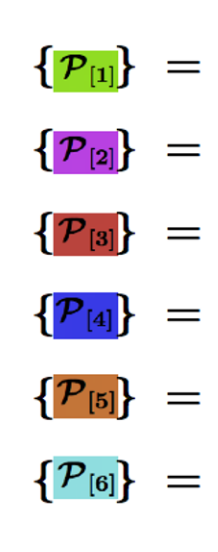}
	\includegraphics[height = 0.3\textwidth]{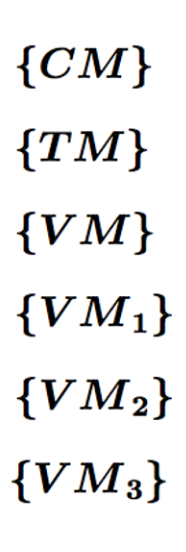}}
	\caption{Left: The S4 permutahedron with 1d, N=4 supermultiplets highlighted in different colors. Right:  Kevin's Pizza- the partition diagram of $\mathbb{S}_4$ into the six supermultiplets, with orbits $*$ generated by transposition.}
	\label{fig:permutahedronpizza}
\end{figure}

We find a set of hopping operators $\bm{\mathcal{H}}_J$ through the following definitions, where we have calculated the permutation matrices from \ref{eqn:Lconvention}:
\begin{gather}
    \bm{\mathcal{P}}_{\hspace{1pt}\widehat{\text{I}}} =  \bm{\mathcal{P}}_{\hspace{1pt}1}\bm{\mathcal{H}}_{\hspace{1pt}\widehat{\text{I}}}\;,
    \hspace{\baselineskip}
    \bm{\mathcal{H}}_{\hspace{1pt}\widehat{\text{I}}} = \bm{\mathcal{P}}_{\hspace{1pt}1}\hspace{-2pt}^\text{T}\;\bm{\mathcal{P}}_{\hspace{1pt}\widehat{\text{I}}}
    \label{eqn:hopperdefinition}
\end{gather}
and we have chosen the convention of acting on the right. Hopping operators can be thought of as generators of the unsigned L-matrices of a multiplet: by acting on the initial node $\bm{\mathcal{P}}_{1}$, we may generate the set of nodes $\bm{\mathcal{P}}_{\hspace{1pt}\widehat{\text{I}}}$. Similarly, one can define a set of left-acting hopping operators and generate the same set of nodes by left-multiplication. Furthermore, Hopping operators are always symmetric--more on this in later sections, so the left-handed Hopping operators are equal to the right-handed Hopping operators of the adjoint set $\bm{\mathcal{P}}_{\hspace{1pt}\widehat{\text{I}}}\hspace{-2pt}^\text{T}$. 

By multiplying a permutation $\bm{\mathcal{M}}$ on the right of $\bm{\mathcal{P}}_{\hspace{1pt}\widehat{\text{I}}}$, we implicitly are permuting $\bm{\mathcal{P}}_{\hspace{1pt}1}$ and performing conjugation on $\bm{\mathcal{H}}_{\hspace{1pt}\widehat{\text{I}}}$. This operation and demonstrates the existence of a conjugacy class of Hopping operators, as well as an automorphism group of permutations for each element of the class.
\begin{align}
\begin{split}
    \bm{\mathcal{P}}_{\hspace{1pt}\widehat{\text{I}}}\;\bm{\mathcal{M}} &= \bm{\mathcal{P}}_{\hspace{1pt}1}\bm{\mathcal{H}}_{\hspace{1pt}\widehat{\text{I}}}\;\bm{\mathcal{M}}\\ &= \bm{\mathcal{P}}_{\hspace{1pt}1}\;\bm{\mathcal{M}}\;(\bm{\mathcal{M}}^{\hspace{-2pt}^\text{T}}\bm{\mathcal{H}}_{\hspace{1pt}\widehat{\text{I}}}\;\bm{\mathcal{M}})\\
    \bm{\mathcal{P}}_{\hspace{1pt}\widehat{\text{I}}}' &= \bm{\mathcal{P}}_{\hspace{1pt}1}'\bm{\mathcal{H}}_{\hspace{1pt}\widehat{\text{I}}}'\label{eqn:hopperpermute}
    \end{split}
\end{align}

Hopping operators may also be considered as: the {\it {minimal length}} paths\footnote{These minimal paths need not be unique.} along the permutahedron which connect the corresponding nodes of a multiplet, or the fundamental set(s) of $\bm{\mathcal{P}}_{\hspace{1pt}\widehat{\text{I}}}$ which contain the identity element and can be permuted to generate all other cosets. By Equation (\ref{eqn:hopperdefinition}), we have chosen to specify the existence of left-cosets which may be navigated by a left-acting permutation.

Now consider the six 1D $\text{N}=4$ supermultiplets shown in Figure \ref{fig:permutahedronpizza}. By generating the conjugacy classes of Hopping operators, it is found that there is only one element in this class; the Hopping operators are an immutable property of N=4 off-shell theories, just as there is only one subset $\{\mathcal{P}_6\}$ which contains the Identity element. This is not surprising because that subset is equal to the Klein 4-Group, and thus has $\mathbb{S}_4$ as its automorphism group. 

The six subsets are further classified as a group of abnormal cosets--cosets which share both left-acting and right-acting Hopping operators \cite{cianciara_cal_2023}. These are:

\begin{align}
    \bm{\mathcal{H}}_1 &=\text{I}\;\;\otimes\;\;\text{I} = ()\;;\\
    \bm{\mathcal{H}}_2 &=\text{I}\;\;\otimes\;\;\sigma^1 = (12)(34)\;;\\
    \bm{\mathcal{H}}_3 &=\sigma^1\;\;\otimes\;\;\text{I} = (23)(12)(34)(23)\;;\\
    \bm{\mathcal{H}}_4 &=\sigma^1\;\;\otimes\;\;\sigma^1= (23)(12)(34)(23)(12)(34).\label{eqn:examplehoppers}
\end{align}

\begin{figure}[ht!]
    \centering
    \includegraphics[scale = 0.25]{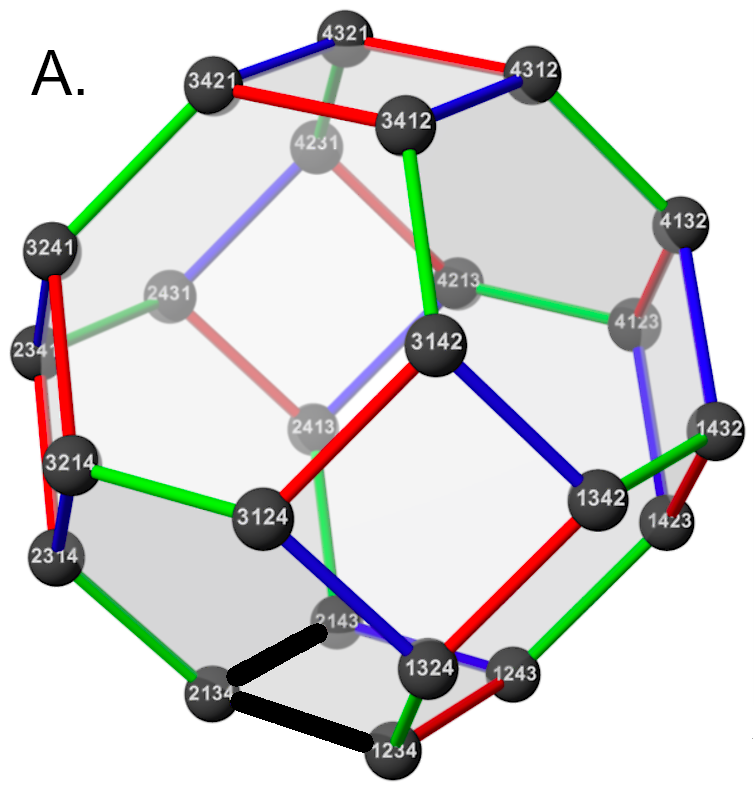}
    \hspace{0.1\textwidth}
    \includegraphics[scale = 0.25]{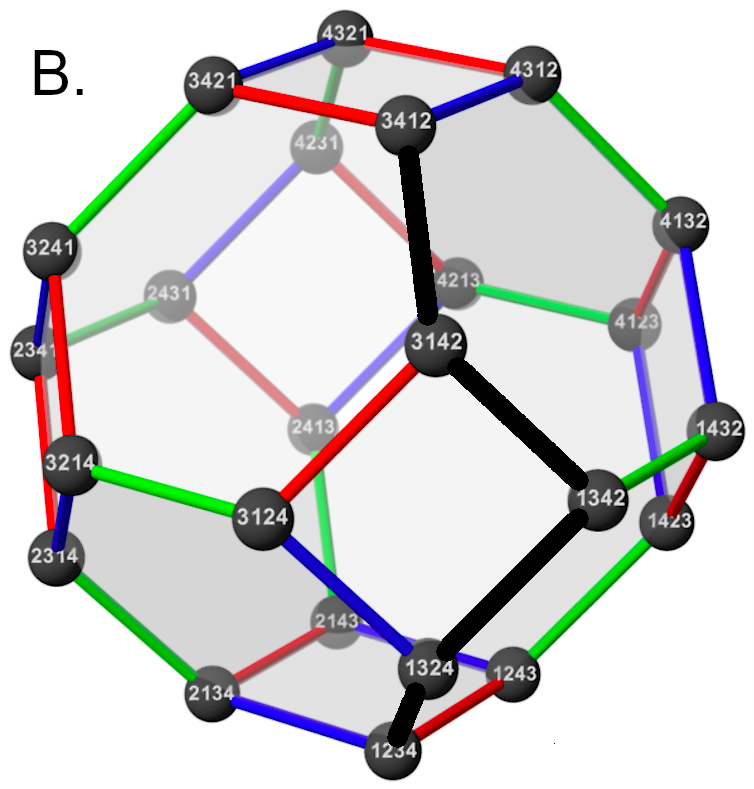}\\
    \includegraphics[scale = 0.25]{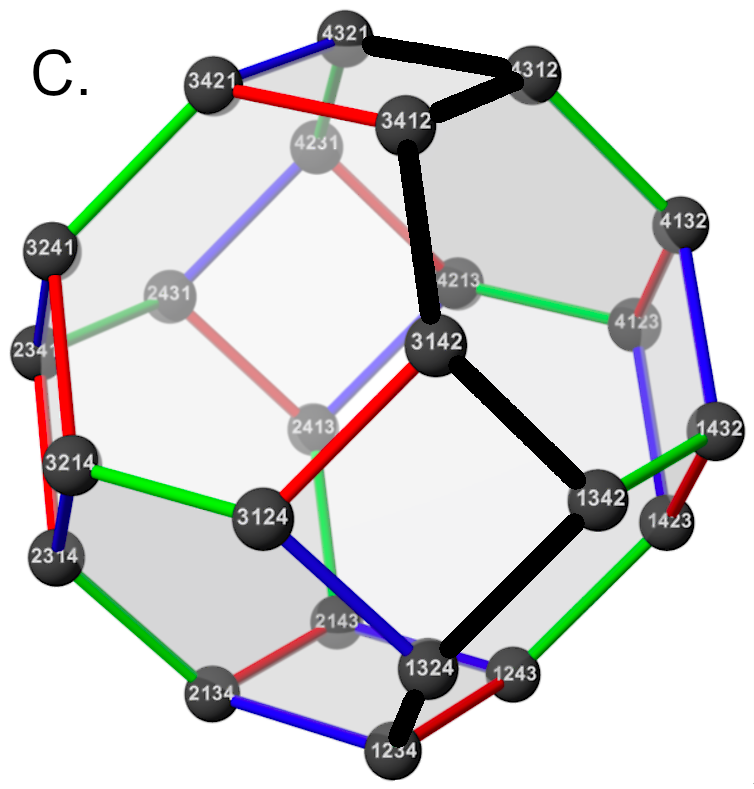}
    \caption{The N=4 Hopping Operators outlined in black as generators of the unsigned L-matrices: A) $\bm{\mathcal{H}}_2$ takes us from the identity element to the element $\langle2143\rangle$, B) $\bm{\mathcal{H}}_3$ takes us from the identity element to the element $\langle4321\rangle$, and C) $\bm{\mathcal{H}}_4$ takes us from the identity element to the element $\langle3412\rangle$. }
    \label{fig:examplehoppers}
\end{figure}

As an example, we show in Figure \ref{fig:examplehoppers} how the hopping operators take us from the identity element $\langle1234\rangle$ at the bottom, to the second $\langle2143\rangle$ third $\langle3421\rangle$ and fourth $\langle4321\rangle$ permutations in the set respectively. Angled brackets are used to denote one-line notation.

From these diagrams, we can understand the symmetries of the $\mathbb{S}_4$ permutahedron and of its subsets without going through a calculation. An investigation into the regularities of these subsets was carried out in \cite{cianciara_300_2021}, where so-called magic numbers became relevant. Magic numbers, denoted by $\mathcal{G}$, are the sum over distances on the permutahedron between one element and all other elements in a subset. Coincidentally for the present case, these distances are also equal to that of the shortest cycle which visits each point in the subset. From its definition, we can say that the value of $\mathcal{G}$ is fully determined by a set of hopping operators, and in this case is always equal to twelve. 

Extending to N=8 takes into consideration the $\mathbb{S}_8$ permutahedron, which is known as the omnitruncated 7-simplex with 40,320 vertices. Using (\ref{eqn:hopperpermute}) in an exhaustive search, we find 30 unique Hopping operator sets, with each of which having an automorphism group of order 1,344. We then searched by left-multiplication for left-cosets, and found for each a total of 5,040 unique cosets. In summary, inhabiting the $\mathbb{S}_8$ permutahedron are 30 unique supermultiplets at every vertex, each of which generate 5,040 left-cosets, totalling 151,200 unique unordered sets of $\bm{\mathcal{P}}_{\hspace{1pt}\widehat{\text{I}}}$. An important change from N=4 is that the set of all hopping operators now forms a nontrivial conjugacy class. 

Furthermore we find, in agreement with \cite{cianciara_cal_2023}, 168 abnormal cosets for each set of hopping operators. We reach this number by considering the 5,040 left cosets generated, and arguing that since the diadems evenly partition the right-handed permutahedron, they must also evenly partition the left-handed permutahedron. Therefore by parsing through the left handed hopping operators, we would expect to find $\frac{5,040}{30} = 168$ which match the right-handed set. We also later revisit the abnormal cosets through a group theoretical approach, which we find inspires continued study.

Using terminology from the aforementioned works, our hopping operators correspond to ``Diadems" which can be used to tile the permutahedron with cosets. Conversely, the quotient of the permutahedron by a single diadem gives the nodes and edges representing the left cosets and their inter-connectivity. For the example of $\text{N}=4$, the quotient leads to a single hexagonal face forming a 6-cycle of the supermultiplets. It follows that one can generate 30 different quotient graphs for N=8, each of which would have 5,040 nodes. It then may be of interest to study the arrangement of the abnormal cosets inside these quotient graphs.

In recent work \cite{cianciara_cal_2023}, it was found that magic numbers of all N=8 supermultiplets are equal to 112, regardless of the distinction of diadems. The authors are motivated to conjecture the ``magic number rule": that all supermultiplets have the same magic number which can be estimated based on the average distance between vertices. We find that a similar investigation in $\text{N}=16$ leads to a counter example of this rule, and additionally that magic numbers are generally not invariant to diadems; it appears to be a special property of supermultiplets in $\text{N}=4$ and $\text{N}=8$ where this has held. As for the compute method for these numbers, the shortest descending path from a given node on the permutahedron to the identity can be traced by a specific implementation of bubble sort \cite{pons_combinatorics_nodate}. This is clear when we consider the inversion number of a given permutation $\bm{\mathcal{P}}$. The shortest path from $\bm{\mathcal{P}}$ to the identity is bounded from below by the inversion number of $\bm{\mathcal{P}}$. After each iteration of the bubble sort, the inversion number is reduced by 1 and the new permutation is therefore closer to the identity element.


\section{Classification of Adinkra Topologies}
\label{sec:topologies}

$~~~~$ So far, the enumeration of elements of a conjugacy class have given us a set of ``diadems" alike; as will be expanded on in the following section, it has been found that there is an indirect connection between diadems and different Adinkra chromotopologies. First introduced by authors C.F. Doran et al. \cite{doran_graph-theoretic_2007}, the ``topology" of an Adinkra refers to the set of vertices and the set of edges which connect them, while ignoring sign factors, colors, height specification, and edge direction. For the purpose of this work, we only consider the valise Adinkras, and consider the Adinkra's ``topology" to be an undirected, N-biregular graph of order d. The ``chromotopology" is the topology of an Adinkra along with the coloring of its edges. It was the subsequent finding that quotienting a colored N-cube by different doubly even codes leads to the derivation of different chromotopologies \cite{doran_relating_2008}. Using doubly even codes, we find the topology of the $\text{N}=16$ Spinning Particle Multiplet with the help of its distance matrix spectra. Furthermore, we expand on the connection to diadems mentioned.

\subsection{A Review of Doubly Even Codes, Quotients, and Families}
\label{ssec:review}

$~~~~$ In the seminal works \cite{doran_graph-theoretic_2007,doran_topology_2008,doran_relating_2008,doran_codes_2011}, the engineering of Adinkras was introduced through an extensive collaborate effort. Since we limit ourselves to valise Adinkra in this work, we should note that the Adinkras we depict are a subset of all possible engineerable Adinkras. The work to classify the many possible height configurations was performed in \cite{doran_graph-theoretic_2007}, by developing Adinkra specific vertex-raising theorems. It is from this work that Adinkra ``engineering" was introduced into the SUSY representation theory toolkit. Through the remaining works, coding theory and various ideas from discrete mathematics provided a means to jump several of the remaining hurdles, which we now expand on in this work.

In \cite{doran_relating_2008}, a derivation of Adinkra topologies up to $\text{N}=16$ is given, as well as a full tabulation of topologies up to $\text{N}=8$ (which is extended up to $\text{N}=11$ in \cite{doran_codes_2011}). It is shown by a derivation using monomial elements of the SUSY operators that an Adinkra representing an off-shell supermultiplet has a fundamental topology of a k-fold iterated $\mathbb{Z}_2$-quotient of an N-cube. The k quotients must be performed such that only like-edges and vertices overlap without loss of color or dashing information. It is shown that the topologies allowed by this condition are bounded by a maximal and a minimal representation, the former being the N-cube and the latter being the $\varkappa$-fold quotient, where $\varkappa$ is given by a recursive formula. The remaining topologies are intermediate; these for $\text{N}\leq8$ have been tabulated and the recursive process for generating all $\text{N}>8$ topologies has been outlined. In analogy to the practical usage of error correcting codes, the act of iterative quotients are similar to the decoding of an incoming error corrected message, where the message itself is the closure of the Garden algebra, and a certain amount of redundancy is associated with the dimensionality of any non-minimal representation.

To perform a quotient, we start with the maximal representation of an N-color Adinkra, which is the N-dimensional Hamming cube with vertices and edges $v_i, e_i\in (\mathbb{Z}_2)^\text{N}$. Doubly even codes are a set of vectors $b_i\in (\mathbb{Z}_2)^\text{N}$ which have hamming weights equal to some multiple of 4. They are presented in the referenced body of work in the form of generator matrices which, by all linear combinations of rows, give the complete list of vectors in the code. A single code with length N and dimension k can be used to perform k quotients on the N-Hamming cube, by identifying all groups of vertices which are equivalent up to some element of $b_i$. This is best understood graphically as the ``orbifold" operation \cite{doran_topology_2008} applied k number of times. By combination of these steps, one simply requires a valid $\text{N}\times \text{k}$ generator matrix in order to properly quotient an N-cube k times. With the choice $\text{k}=\varkappa$, we must use a code which quotients the maximum number of times without destroying the off-shell closure, which is therefore called a maximal doubly even code. Furthermore, any code generating matrix has a permutation equivalence class, which can be obtained by permuting the columns of the generator. These permutation equivalence classes are each descriptors of a different Adinkra topology, and since columns of the matrix represent the N different colors, elements within these classes each correspond to different chromotopologies.

The classification of Adinkra topologies is done by the nature of their codes, and can be denoted by different families of Coxeter groups, such as $\text{I}_\text{N}$, $\text{A}_\text{n}$, $\text{D}_\text{n}$, $\text{E}_7$, and $\text{E}_8$. In \cite{doran_codes_2011}, the number of permutation-equivalent codes for each topology are tabulated, and there exist 30 such codes for the minimal representation of $N=8$, corresponding with our count of 30 different sets of hopping operators. The number and order of automorphism groups of codes are useful knowledge in novel applications of coding theory and error correction, and have been a topic of investigation long preceding Adinkras \cite{conway_binary_1992}. Given a one-to-one correspondence between binary self-dual doubly even codes and Adinkra topologies, as well as codes being a fundamentally more efficient representation than lists of matrices, it is clear that they are necessary in the study of 1D $N\geq16$ SUSY.

Now given the necessary tools, we may examine $N=16$ off-shell theories through a fine lens and are much better equipped to understand the off-shell problem. From this point we refer to these methods described, including the classification of topologies by codes, as the ``engineering principles" of Adinkras. We should note that while there is certainly a direct link between Adinkra topologies and doubly even codes, it is still unproven as to whether the different permutation equivalence classes of codes can be considered a complete descriptor of all possible Adinkra topologies. Fortunately, the Adinkra considered herein is shown to be ``well engineered".

\subsection{``Decoding" the \texorpdfstring{N$=16$}{TEXT} Spinning Particle Multiplet}
\label{ssec:decoding}

\begin{figure}[ht!]
    \centering
    \includegraphics[scale = .2]{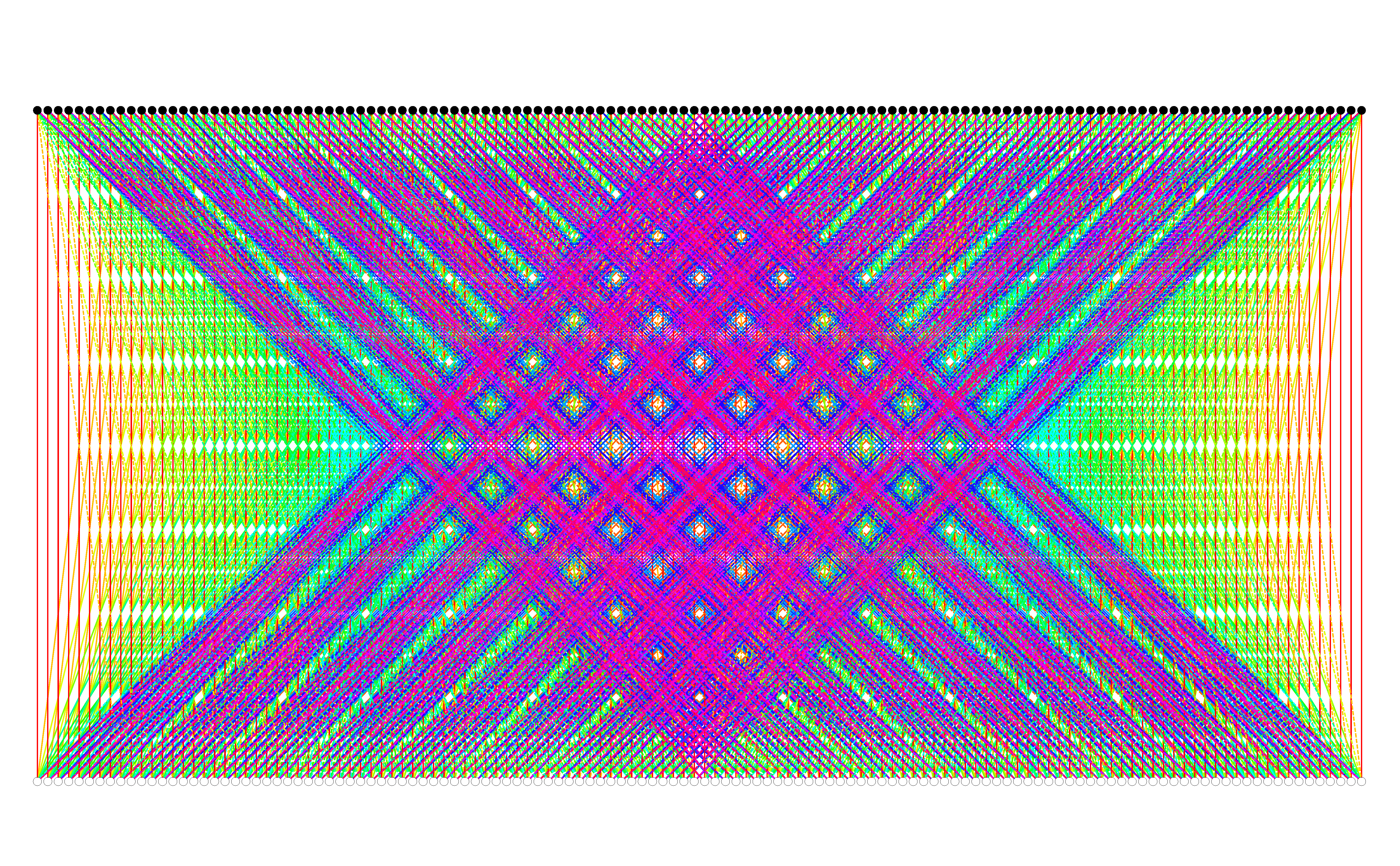}
    \caption{The Adinkra for the N=16 Spinning Particle theory: white nodes and black nodes represent 128 Bosons (bottom) and 128 Fermions (top), and 16 colors represent 16 different supersymmetry transformations.}
    \label{fig:spinning}
\end{figure}

$~~~~$ The Spinning Particle supermultiplet couples a 1D minimal supergravity with a 1D N-extended matter supermultiplet \cite{gates_theory_1995,gates_theory_1996}. In this referenced work, rules for representing arbitrary N-extended Spinning Particle theories were introduced, which is reduced to finding a set of matrices $\textbf{L}_I$ and $\textbf{R}_I$ with $I=1\dots \text{N}$ which satisfies the constraint of antisymmetry and the Clifford algebra (\ref{eqn:grdn2}). It was then shown that these matrices can be found via a recursive formula using Pauli matrices. We have reproduced the L-matrices for $\text{N}=16$ in Appendix \ref{app:ExplicitLMatrices}, where we have used the since-updated convention $\textbf{R}_I = \textbf{L}_I{}^{\text{T}}$, as well as the convention of weak Bruhat ordering of permutations. These matrices represent the transformation rules for the 128 superpartners included in the multiplet, and allow us to construct the Adinkra in Figure \ref{fig:spinning}.   

It is also noted that there should be a number of inequivalent representations $\textbf{L}_I$ \cite{gates_theory_1995}, each corresponding to a unique supermultiplet. For $N=8$, we have discussed that there are $30*5,040$ inequivalent representations of $|\textbf{L}_I|$ which satisfy the Garden Algebra alone. To calculate the number of choices for sign factors, we can use the general formula \cite{gates_n4_2016}: 
\begin{gather}
	2^{2d-1}\cdot|C|\;;\label{eqn:numsigns}
\end{gather}
\noindent
where $|C|$ is the number of elements in the code $C$ of the representation. There are 524,288 sign factor choices $\bm{\mathcal{S}}_{I}$ per set of permutations $\bm{\mathcal{P}}_{I}$, leading to a total of 79,272,345,600 choices of $\textbf{L}_I$. Given the condition of antisymmetry, we can find the number of permutation sets by considering only symmetric matrices, or equivalently by counting the total number of involutions in $\mathbb{S}_8$ which is 764. We have additionally verified that the set of involutions acting on each set of Hopping operators produces no duplicates up to a reordering of colors. There are thus $30*764 = 22,920$ choices of permutation matrices for the $N=8$ spinning particle multiplet. We may find the sign factors by a similar argument; due to the restriction of antisymmetry, the aforementioned formula should be modified to:
\begin{gather}
	2^{d}\cdot|C|\;;\label{eqn:numsignsAS}
\end{gather} 
\noindent
which is proven by the following: consider a set of matrices $\textbf{L}_I$ which satisfy both the Garden algebra and antisymmetry; these will have associated sets of sign factor matrices $\Sigma^{\text{L}}$ and $\Sigma^{\text{R}}$, whereby an augmentation of the form 
\begin{gather}
	\textbf{L}_I{}' = \Sigma^{\text{L}}\textbf{L}_I\Sigma^{\text{R}}\;\label{eqn:augmentation}
\end{gather}
\noindent
will leave the Garden algebra, by definition, unchanged. To preserve antisymmetry, we require that when flipping the sign of row k in matrix $I$, we also flip the sign of column k. This requirement states  $\Sigma^{\text{L}} = \Sigma^{\text{R}} = \Sigma$. We then reason that there are $2\cdot2^{d-1}$ unique choices of $\Sigma$, where an additional factor of two is necessary to account for the overall sign of the set. Similar to Equation (\ref{eqn:numsigns}), we must factor in the number of elements in the code $C$ to find the total number of sign factor choices, and so arriving at (\ref{eqn:numsignsAS}). We find that there are 4,096 choices of sign factors, leading to 93,880,320 possible antisymmetric sets $\textbf{L}_I$. 

As an important note, it has been proven that the number of possible $\mathbb{Z}_2$ quotients determines the dimensions of minimal representations, but this was not always known. Based on the observed rules for supersymmetry algebras, the dimensionality of minimal representations grows combinatorically with N. For $\text{N}=8$, the minimal dimension is $\text{d}=8$, so the minimal representation is a fully connected bipartite graph, therefore only having one possible topology. For $\text{N}>8$, the dimension d grows greater than N, so we cannot assume there is only one possible topology. For theories of $N\leq16$, the topologies of minimal representations have previously been derived and tabulated \cite{doran_relating_2008}. We now know that there are in fact two distinct topologies allowed by minimal $N=16$ off-shell Adinkras, which are characterized by two distinct maximal doubly even codes of length 16 and dimension 8. These codes are reproduced below in (\ref{codesN=16}):

It is clear that such an analysis for determining the total number of inequivalent representations should be effectively the same for $\text{N}=16$ as it was for $N=8$, aside from there being two distinct topologies. 
\begin{align}
    \text{D}_{16}: \hspace{\baselineskip}\begin{tabular}{c}
        0000000000001111\\
        0000000000111100\\
        0000000011110000\\
        0000001111000000\\
        0000111100000000\\
        0011110000000000\\
        1111000000000000\\
        0101010101010101
    \end{tabular}\hspace{2\baselineskip}
    \text{E}_{16}: \hspace{\baselineskip}\begin{tabular}{c}
         00000000 00001111\\
         00000000 00111100\\
         00000000 11110000\\
         00000000 01010101\\
         00001111 00000000\\
         00111100 00000000\\
         11110000 00000000\\
         01010101 00000000
    \end{tabular}
    \label{codesN=16}
\end{align}

To understand the spinning particle multiplet's topology, we base our analysis on derived Hopping operators from codes. The Hopping operators of the $\text{N}=16$ Spinning particle multiplet are equivalent to its permutation matrices $\bm{\mathcal{P}}_{I}$, since the identity element is included by construction. These matrices are given explicitly in Appendix \ref{app:ExplicitHoppers}.

A \textit{Mathematica} function was designed for the generation and analysis of the Hopping operators associated with both codes listed above. The function takes as input a code generator matrix of any length N and dimension k, and outputs the result of the quotient in the form of unsigned matrices $|\bm{\Gamma}_\text{I}|$ as in Equation (\ref{eqn:grdn2}). Given the two newly derived hopping operators as well as those of the Spinning particle multiplet, we calculated the adjacency and distance matrix of each, as well as their spectra. Other graph properties, like radius, diameter, and girth are the same for each and are all equal to 4, indicating that any vertex is either a distance of 2 or 4 away from any other vertex at the same height. The eigenvalues of the adjacency and distance matrices for the two derived topologies are compared in Table \ref{tab:spectra}.

\begin{table}[htp!]
    \centering
    \begin{tabular}{|c|c|c|c|c|c|}
    \hline
     & Adj. & Dist. & & Adj. & Dist. \\ 
    \cline{2-3}\cline{5-6}
     $\;\;$\makecell{$\text{D}_{16}$\\$\;$}$\;\;$ & \makecell{-16 \\ 16 \\ -8 $\color{gray}(\times 28)$ \\ 8  $\color{gray}(\times 28)$\\ 0  $\color{gray}(\times198)$} &\makecell{746 \\ 42 \\ -38 $\color{gray}(\times 28)$\\  10  $\color{gray}(\times 70)$\\  -6  $\color{gray}(\times 28)$\\  -2 $\color{gray}(\times 128)$} &  $\;\;$\makecell{$\text{E}_{16}$\\$\;$}$\;\;$ & \makecell{-16 \\ 16 \\ -8 $\color{gray}(\times 28)$\\ 8 $\color{gray}(\times 28)$\\ 0 $\color{gray}(\times 198)$} & \makecell{ 704 \\ 96 \\ 96 \\ -32 $\color{gray}(\times 28)$\\ 0 $\color{gray}(\times 225)$}\\\hline
    \end{tabular}
    \caption{Adjacency and distance spectra of the topologies of the two Adinkras from (\ref{codesN=16}).}
    \label{tab:spectra}
\end{table}

The Spinning Particle Adinkra is found to have identical spectra to those of the $\text{E}_{16}$ Adinkra, so we classify it as having such a topology, assuming there are no co-spectral mates which also have eluded the derivation of the codes. These two Adinkras have different diadems, which is not surprising given the assumedly large automorphism group. In addition to calculating the spectra, we have calculated the magic numbers (denoted by $\mathcal{G}$) using the algorithm mentioned in Section \ref{sec:hoppers}. These numbers are all found to be different:
\begin{align*}
    \mathcal{G}^{\text{D}_{16}} &= 32,768\;; \\
    \mathcal{G}^{ \text{E}_{16}} &= 32,512\;; \\
    \mathcal{G}^{\text{SP}} &= 48,896\;.
\end{align*}

In summary, we have confirmed that the 1D, $N=16$ Spinning particle multiplet has the topology of $\text{E}_{16}$, better known as $\text{E}_8 \times \text{E}_8$, as opposed to the alternative $\text{D}_{16}$, better known as $\text{SO}(32)$; the gauge symmetry groups of the Heterotic strings. We have demonstrated that codes can be used to classify topologies in practice, using a platform like \textit{Mathematica}. We also report that the topologies have different magic numbers, as does the spinning particle multiplet which shares a topology, counter to the conjecture made previously in \cite{cianciara_cal_2023}. We note that the distance spectrum is invariant to diadems, and is not invariant to topology, thus acting as a signal of topological equivalence sufficient for the needs of this work. Although, we haven't proven that the distance spectrum is injective over the domain of all Adinkra topologies, we conjecture that the distance spectrum is a deterministic description of graph topology for all Adinkras with N $\leq16$.

There are several important takeaways from these results. First, we can now classify supermultiplets by their topology without suffering from problems of high dimensionality, and as we will note, this process can be done directly. It follows that we can create arbitrary off shell supermultiplets given a doubly even code and a properly colored and dashed N-Cube. In the ``Genomics" spirit, we have stepped forward in sequencing known $N=16$ theories by analyzing their graph invariants. 

\subsection{Chromotopologies in \texorpdfstring{$(\mathbb{Z}_2)^{N}$}{TEXT}}
\label{ssec:chromotopologies}

We now will expand on the following statement: different chromotopologies of Adinkras can be derived directly from the colored N-cube. The N-Cube Adinkra is assigned its colors based on the dimensions of the N-Hamming cube. First, we start with the labeled Hamming cube of N-bits, and label the edges of this graph by the single bit-flips which generate each vertex. The set of N single bit-flips forms an orthogonal basis. We can then color the edges of the resulting graph, where the different  basis vectors $\bm{a}^{(\text{I})}$ map to the different colors $\text{I}\in 1\dots N$.

If we then calculate the different colored adjacency matrices, we find that they may be represented as Kronecker products of elements of $\mathbb{S}_2$, denotable as:
\begin{align}
    \mathcal{A}^{(\text{I})} &= (\sigma_1)^{a^{(\text{I})}_1}\otimes(\sigma_1)^{a^{(\text{I})}_2}\otimes\dots\otimes (\sigma_1)^{a^{(\text{I})}_N} \label{eqn:Ncubehoppers}.
\end{align}
\noindent
The Hopping operators of the N-cube Adinkra are found by permuting the vertices such that the 1st bit in each basis vector is 1, giving us the form of $\Gamma_I$ in (\ref{eqn:grdn2}). In this work, the \textit{Mathematica} function \texttt{FindIndependentVertexSet} was used to partition vertices, since it will fail to run if there doesn't exist such a set.

When we quotient the N-cube by a code $C$, we move to a new basis $\bm{a}^{\prime(\text{I})}$ of lower dimension. We can assume that any Adinkra obtained by a quotient of the N-cube has matrices $\mathcal{A}^{(\text{I})}$ which can be represented in the form of \ref{eqn:Ncubehoppers}, or are equivalent up to a permutation of labels to such a representation. The reason for this is inherent in the operation we are performing; since we are dividing a generator of codes (the set of all basis vectors in the N-cube) by a generator of codes (a set of linear combinations of those basis vectors), we must be able to obtain a generator of codes as a result. This is a result of the quotient lattice remaining cubical. 

But do all quotients of codes result in a vector-like form \ref{eqn:Ncubehoppers}? This question has investigated in the attempt to connect concepts previously explored in \cite{gates_n4_2016} with the more recent ideas in polytopic representation. For a maximal doubly even code of an $\text{N}=8$ representation, it was reported in \cite{doran_codes_2011} that there are 30 permutation-equivalent choices which are unique. This number is suspicious since it is the same as the number of unique Hopping operators for minimal $\text{N}=8$ theories. An effort was put forth to calculate all 30 Hopping operator sets from the permutation-equivalent codes, and it was found that the two numbers do not form a correspondence. In fact, the 30 new Hopping operators were found to all be different ordered versions of the same set, thus each belonging to the same diadem. Previously, this diadem, which is equivalent to $(\mathbb{S}_2)^3$ and as such can be described by the form \ref{eqn:Ncubehoppers}, was termed the ‘Diadem(8)’ Octet in \cite{bristow_note_2022}. The derivation of the different diadems for $\text{N}=8$ was performed previously by a counting argument involving the group of affine transformations acting on the vertices of a 3-cube \cite{gates_n4_2016}. It is possible that although the different codes do not directly give us the 30 unique sets of Hopping operators, the same counting argument could be sufficient for both the diadems and the chromotopologies of $\text{N}=8$. 

It follows that the new vectors $\bm{a}^{\prime(\text{I})}$ are themselves a code which can be used inversely to recover $C$. Equivalently, we may demand that the hopping operators of an off-shell theory are able to derive a doubly even code from the N-cube. A less obvious discovery is that when considering minimal representations for $\text{N}=0 \text{ mod } 8$, the code found from the hopping operators is permutation-equivalent to the code divided by in the quotient.

For theories of larger N, we find that there are a growing number of hopping operators in the subset $(\mathbb{S}_{2})^n \in \mathbb{S}_d$. For $\text{N}=16$, there are an unknown number of these hopping operators, which may be investigated using the General Linear group as the automorphism group of $(\mathbb{Z}_2)^n$.

To understand how we might enumerate the hopping operators of $\text{N}=16$, consider the structure of $\text{N}=8$. The single set of hopping operators has an abnormal coset group of order 168: this is precisely the order of $\mathbb{GL}((\mathbb{F}_2)^3,2)$, which we denote here as $\mathbb{GL}(3,2)$. We also may deduce that all hopping operators in $\text{N}=8$ have a set of abnormal cosets of the same order, because as sets they are equivalent up to left and right permutation. Noting that $\mathbb{GL}(3,2)$ is the automorphism group of $(\mathbb{Z}_2)^3$, it may also be represented as the set of matrices $\text{S}_{\text{aut}}\subset \mathbb{S}_8$ over which $(\mathbb{S}_2)^3$ forms a conjugacy class. From further investigation it was found that there are 8 matrices in $\text{S}_{\text{aut}}$ for every element of $\mathbb{GL}(3,2)$, and that these matrices show an 8:1 correspondence to the elements, where each of the 8 matrices are equivalent up to permutation by an element in $(\mathbb{Z}_2)^3$. This is logical because any conjugation of $(\mathbb{Z}_2)^3$ by an element of itself is an order-preserving isomorphism of $(\mathbb{Z}_2)^3$. It follows that the hopping operators can be derived directly, as: $\mathcal{O}(\mathbb{S}_8) \div \mathcal{O}(\mathbb{GL}(3,2)) \div \mathcal{O}(\mathbb{S}_2{}^3)  = 30$. This result is previously known from the standpoint of Adinkras as graphs, and is stated in \cite{gates_n4_2016}. 

By exploiting the action of the General Linear group on $(\mathbb{Z}_2)^7$, it may be possible to extend this method to $N=16$. The main hurdle for this is that $ \mathcal{O}(\mathbb{GL}(7,2))$ is no longer equal to the order of the abnormal coset group. The order of the automorphism group for a subset of $(\mathbb{Z}_2)^7$ is the value of importance. A preliminary calculation was done to find the order of the group automorphism on all 35 combinations of 4 elements  $\text{Z}_{(4)}\subset (\mathbb{Z}_2)^3$ which include the identity. We found that the different subsets have different automorphism groups, leading to two possible values of $\mathcal{O}(\text{Aut}(\text{Z}_{(4)}))$: 6 and 24. The values 24 were found only in the subsets which can be read as an even code of length 4 and dimension 3, of which there are 7. It is worth pointing out that the orders 6 and 24 are the orders of the $\mathbb{S}_3$ and $\mathbb{S}_4$ groups, or $\mathbb{GL}(2,2)$ and $\mathbb{PGL}(2,3)$ respectively, and that we may want to be familiar with the underlying subspaces of $(\mathbb{Z}_2)^3$. If we have a selection of $\textbf{I}\;+ $ any 3 elements of $(\mathbb{Z}_2)^3$, we will have an automorphism group of at least order 6 because there will be 3! ways to recover the same set, corresponding to permutations of the elements of the set which are not the identity. If we instead limit ourselves to even codes, one dimension can be consolidated and the cardinality of our group can be reduced by half. Any choice can therefore be mapped to $(\mathbb{Z}_2)^2 \cong (\mathbb{S}_2)^2$ which has an automorphism group $\mathbb{S}_4$. From this we can attempt to understand the origin of the different automorphism groups, and results imply that the abnormal coset groups have different orders depending on the reducibility of their Hopping operators, and are therefore not diadem-invariant. 

This calculation has posed the question: how can we determine the numbers $\mathcal{O}(\text{S}_{aut})$ for the two distinct $N=16$ topologies, without having to fully grasp $\mathbb{GL}(2,7)$? This would indeed require an understanding of the automorphism groups of different orders which manifest for subsets of $(\mathbb{Z}_2)^{7}$, which has yet to be contemplated.

\section{Review of Some On-Shell Theories}
\label{sec:maxwell}

$~~~~$ The purpose of this section is to introduce the subjects of newly developed methods in this work. Four well-known supermultiplets are presented, reduced to 0-brane and realized in the form of Adinkras. We first present the 4D, $\mathcal{N} = 4$ abelian Vector supermultiplet and the 10D, $\mathcal{N} = 1$ abelian Vector supermultiplet each without any auxiliary fields. We then present the 4D, $\mathcal{N} = 4$ Maxwell supermultiplet and the 4D, $\mathcal{N} = 4$ Vector-Tensor supermultiplet. Given the standpoint of Adinkras in this work, we forego a display of the Lagrangian and other analytical figures which have previously been explored. We also make note that the 0-brane reduction of a theory is taken to be a reliable representation of the 4D theory from the standpoint of supersymmetry dynamics \cite{calkins_is_2014}.

The methods mentioned are contained in the form of a \textit{Mathematica} package. Specialized functions were developed to extract L and R matrices, generate Adinkras, and compute Hopping operators from the field component notation. Included in the functionality of this package is the calculation of the holoraumy, HYMNS, magic numbers, and the aforementioned N-cube quotienting function. It has also been a large part of this effort to generate friendly representations of large L and R-matrices. All displays of matrices in this work were generated by functions which attempt to find the most efficient display form, including those which are products of Pauli matrices.   

\subsection{Review of 10D, 4D Abelian Vector Supermultiplets}

$~~~~$ The 10D, 6D, and 4D abelian SUSY-YM theories, referred to as abelian Vector supermultiplets, are related by dimensional reduction first in \cite{brink_supersymmetric_1977} and further explored through a calculation of their holoraumy in \cite{gates_infinite-dimensional_2022}. 

The 10D, $\mathcal{N}=1$ abelian Vector supermultiplet is composed of a single 10D vector field and a Majorana fermion. The theory has a closed supersymmetric algebra under equations of motion, and it is fully on-shell with no off-shell degrees of supersymmetry. Using the 10D gamma conventions in \cite{gates_infinite-dimensional_2022}, the transformation rules are:
\begin{align}
	D_aA_\mu &= (\sigma_\mu)_{ab}\lambda^b \\
	D_a\lambda^b &= i(\sigma^{\mu\nu})_a{}^b\partial_\mu A_\nu \;.
	\label{eqn:10Dabelianvector}
\end{align}
\noindent
Given by the procedures of 0-brane reduction and matrix representation outlined in \cite{gates_jr_4d_2009}, we present the L-matrices in Appendix \ref{app:ExplicitLMatrices} and the corresponding Adinkra shown below. There are 9 boson 16 fermion field components in total, where we have made gauge choices to ignore $A_0$
\begin{figure}[ht!]
	\centering
	\includegraphics[width = 0.6\textwidth]{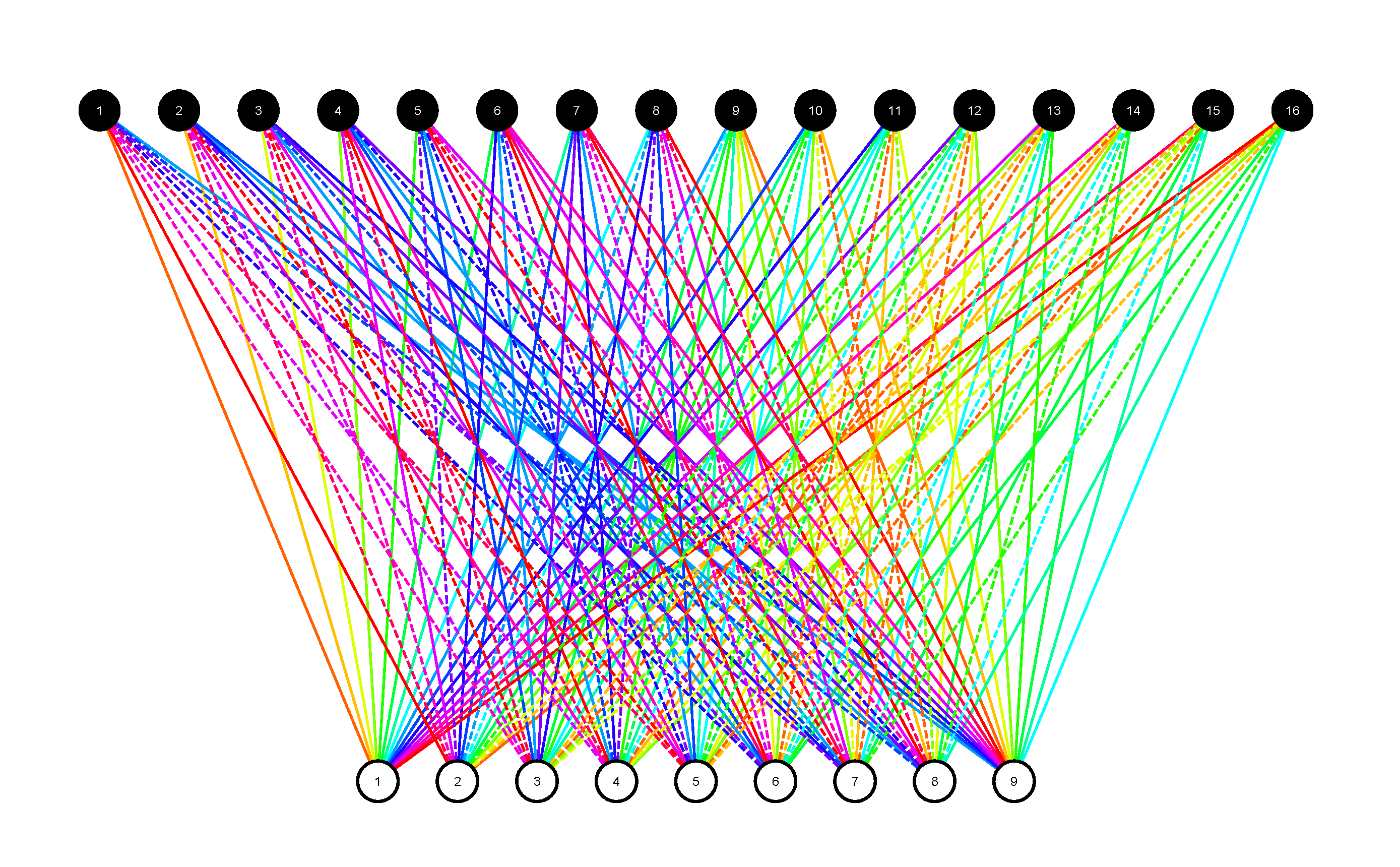}
	\caption{Adinkra for the 10D, $\mathcal{N}=1$ Abelian Vector supermultiplet: consists of 9 Boson (White) and 16 Fermion (Black) field components, and N $=16$ edge colors.}
	\label{fig:10Dabelianvector}
\end{figure}

The 4D $\mathcal{N}=4$ abelian vector supermultiplet is the 4-extended version of the 4D abelian vector multiplet as in \cite{gates_infinite-dimensional_2022}, the dimensional reduction of the theory above. It can be described as one $\mathcal{N}=1$ on-shell 4D vector multiplet coupled with a triplet of $\mathcal{N}=1$ on-shell 4D chiral multiplets. The vector multiplet has a vector $A_\mu$ and a majorana fermion $\lambda_a$, while the chiral multiplets have each a scalar field $A$, a psuedoscalar field $B$, and a majorana fermion $\psi_a$. The theory has a closed supersymmetric algebra under equations of motion, and it is also fully on-shell. The transformation rules are given as:
\begin{align}
    D_aA^{\mathcal{I}} &= \psi_a^{\mathcal{I}}\\
    D_aB^{\mathcal{I}} &= i(\gamma^5)_a{}^b\psi_b^{\mathcal{I}}\\
    D_aA_\mu &= (\gamma_\mu)_a{}^b\lambda_b\\
    D_a^{\mathcal{I}}A_\mu &= -(\gamma_\mu)_a{}^b\psi_b^{\mathcal{I}}\\
    D_a^{\mathcal{I}} A^{\mathcal{J}} &= \delta^{\mathcal{I}\mathcal{J}}\lambda_a - \epsilon^{\mathcal{I}\mathcal{J}}{}_{\mathcal{K}}\psi_a^{\mathcal{K}}\\
    D_a^{\mathcal{I}}B^{\mathcal{J}} &= i(\gamma^5)_a{}^b[\delta^{\mathcal{I}\mathcal{J}}\lambda_b + \epsilon^{\mathcal{I}\mathcal{J}}{}_{\mathcal{K}}\psi_b^{\mathcal{K}}]\\
    D_a\psi_b^{\mathcal{I}} &= i(\gamma^\mu)_{ab}(\partial_\mu A^{\mathcal{I}}) - (\gamma^5\gamma^\mu)_{ab}(\partial_\mu B^{\mathcal{I}})\\
    D_a\lambda_b &= -i\frac14([\gamma^\mu\;,\;\gamma^\nu])_{ab}(\partial_\mu A_\nu - \partial_\nu A_\mu)\\
    D_a^{\mathcal{I}}\lambda_b &= i(\gamma^\mu)_{ab}(\partial_\mu A^{\mathcal{I}}) - (\gamma^5\gamma^\mu)_{ab}(\partial_\mu B^{\mathcal{I}})\\
    D_a^{\mathcal{I}}\psi_b^{\mathcal{J}} &= \delta^{\mathcal{IJ}}i\frac14([\gamma^\mu\,,\,\gamma^\nu])_{ab}(\partial_\mu A_\nu - \partial_\nu A_\mu) + \epsilon^{\mathcal{IJ}}{}_{\mathcal{K}}[i(\gamma^\mu)_{ab}(\partial_\mu A^{\mathcal{K}}) + (\gamma^5\gamma^\mu)_{ab}(\partial_\mu B^{\mathcal{K}})]
\end{align}
\noindent
Given by the same procedures of 0-brane reduction and matrix representation, we present the L-matrices in Appendix \ref{app:ExplicitLMatrices} and the corresponding Adinkra shown below. There are again 9 boson and 16 fermion field components following the reduction and gauging away of $A_0$. 

\begin{figure}[ht!]
    \centering
    \includegraphics[width = 0.6\textwidth]{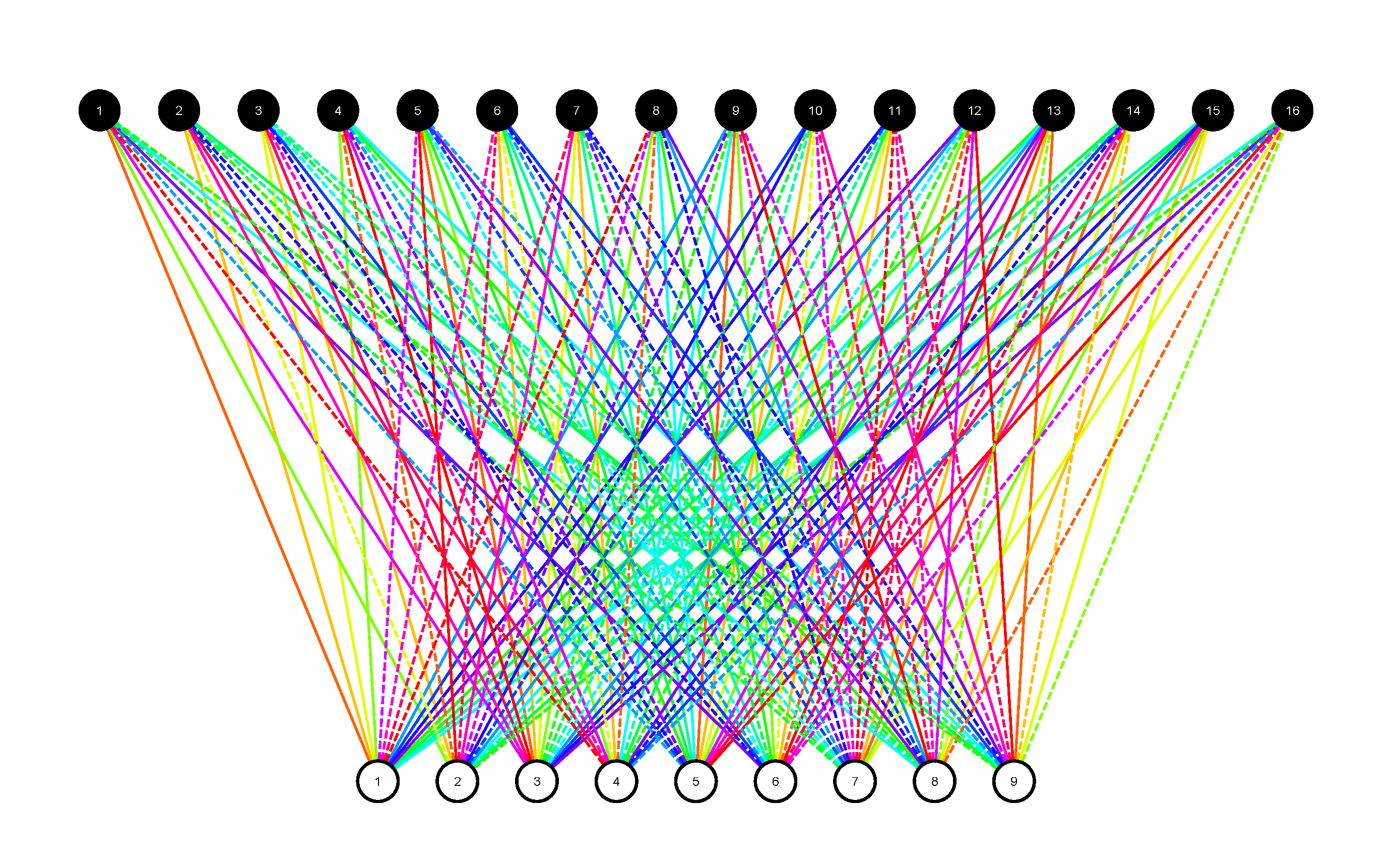}
    \caption{Adinkra for the 4D, $\mathcal{N}=4$ abelian Vector supermultiplet: similar to the previous Adinkra, consists of 9 Boson (White) and 16 Fermion (Black) field components, and N $=16$ edge colors. It should be noted that Adinkras are representations of 1D theories which are unique, yet reliably capture the properties of the higher dimensional theory.}
    \label{fig:vector4Adinkra}
\end{figure}

We are now able to use the full representation toolkit to analyze these theories using their 1-dimensional analogues. Since the Adinkras are both of the same field dimensions, and since they are both 16-biregular, we can say with certainty that they have the same topology. Without considering the structure of the equations of motion and non-closure terms, we can't assume that they are equivalent cosets, because no such classification of on-shell theories has been theorized from Adinkras alone. In this work, we report the existence of more than one pair of permutations $\{\bm{C}^{l},\;\bm{C}^{r}\}$ which takes us from one theory's permutation matrices to the other by the following:
\begin{align*}
\bm{C}^{l}|\textbf{L}_{\widehat{\text{I}}}^{4D}| \hspace{1pt}\bm{C}^{r} \in \{|\textbf{L}^{10D}|\}\;.
\end{align*}
\noindent
These pairs were found using a semi-randomized search, and let us conclude that the unsigned L matrices of the two theories are equivalent up to a renaming of the bosons and fermions. To additionally show an equivalence between L-matrices, there must also be a valid redefinition of the sign factors following these permutations, of which there are many possibilities. 

\subsection{Review of the 4D \texorpdfstring{$\mathcal{N}=4$}{TEXT} Maxwell Supermultiplet}

$~~~~$The 4D $\mathcal{N}=4$ Maxwell Supermultiplet can be described as one off-shell 4D vector multiplet coupled with a triplet of off-shell 4D Chiral multiplets, and has been given previously in matrix representation \cite{calkins_is_2014}. The vector multiplet contains an additional pseudoscalar auxiliary field $d$, while the chiral multiplets contain additional auxiliary scalars $F$ and psuedoscalars $G$, when compared to the 4D abelian Vector supermultiplet. Previously, the valise-Adinkraic representation of this theory was given as well as the Adinkra matrices and the supercommutator algebra results on the 0-Brane \cite{gates_extended_2015}. In both references, the off-shell problem was posited, questioning the existence of an off-shell ``hologram" representation which contains this multiplet embedding. It was shown by example in the former work, that non-closure terms of the 1D Garden Algebra can be related to those of the 4D supercommutator algebra; this inspired the development of an algorithm which completes the closure of the algebra given the L and R-matrices of an on-shell representation \cite{calkins_think_2015}. The SUSY transformation operators have the following rules:
\begin{align}
    D_aA^{\mathcal{I}} &= \psi_a^{\mathcal{I}}\\
    D_aB^{\mathcal{I}} &= i(\gamma^5)_a{}^b\psi_b^{\mathcal{I}}\\
    D_aF^{\mathcal{I}} &= (\gamma^\mu)_a{}^b(\partial_\mu\psi_b^{\mathcal{I}})\\
    D_aG^{\mathcal{I}} &= i(\gamma^5\gamma^\mu)_a{}^b(\partial_\mu\psi_b^{\mathcal{I}})\\
    D_aA_\mu &= (\gamma_\mu)_a{}^b\lambda_b\\
    D_ad &= i(\gamma^5\gamma^\mu)_a{}^b(\partial_\mu\lambda_b)\\
    D_a^{\mathcal{I}}A_\mu &= -(\gamma_\mu)_a{}^b\psi_b^{\mathcal{I}}\\
    D_a^{\mathcal{I}}d &= i(\gamma^5\gamma^\mu)_a{}^b(\partial_\mu\psi_b^{\mathcal{I}})\\
    D_a^{\mathcal{I}} A^{\mathcal{J}} &= \delta^{\mathcal{I}\mathcal{J}}\lambda_a - \epsilon^{\mathcal{I}\mathcal{J}}{}_{\mathcal{K}}\psi_a^{\mathcal{K}}\\
    D_a^{\mathcal{I}}B^{\mathcal{J}} &= i(\gamma^5)_a{}^b[\delta^{\mathcal{I}\mathcal{J}}\lambda_b + \epsilon^{\mathcal{I}\mathcal{J}}{}_{\mathcal{K}}\psi_b^{\mathcal{K}}]\\
    D_a^{\mathcal{I}}F^{\mathcal{J}} &= (\gamma^\mu)_a{}^b\partial_\mu[\delta^{\mathcal{I}\mathcal{J}}\lambda_b - \epsilon^{\mathcal{I}\mathcal{J}}{}_{\mathcal{K}}\psi_b^{\mathcal{K}}]\\
    D_a^{\mathcal{I}}G^{\mathcal{J}} &= i(\gamma^5\gamma^\mu)_a{}^b\partial_\mu[-\delta^{\mathcal{I}\mathcal{J}}\lambda_b + \epsilon^{\mathcal{I}\mathcal{J}}{}_{\mathcal{K}}\psi_b^{\mathcal{K}}] \label{eqn:vectorvaliserules1}
\end{align}

\begin{align}
    D_a\psi_b^{\mathcal{I}} &= i(\gamma^\mu)_{ab}(\partial_\mu A^{\mathcal{I}}) - (\gamma^5\gamma^\mu)_{ab}(\partial_\mu B^{\mathcal{I}}) - iC_{ab}F^{\mathcal{I}} - (\gamma^5)_{ab}G^{\mathcal{I}}\\
    D_a\lambda_b &= -i\frac14([\gamma^\mu\;,\;\gamma^\nu])_{ab}(\partial_\mu A_\nu - \partial_\nu A_\mu)+ (\gamma^5)_{ab}d\\
    D_a^{\mathcal{I}}\lambda_b &= i(\gamma^\mu)_{ab}(\partial_\mu A^{\mathcal{I}}) - (\gamma^5\gamma^\mu)_{ab}(\partial_\mu B^{\mathcal{I}}) - iC_{ab}F^{\mathcal{I}} - (\gamma^5)_{ab}G^{\mathcal{I}}\\
    D_a^{\mathcal{I}}\psi_b^{\mathcal{J}} &= \delta^{\mathcal{IJ}}[i\frac14([\gamma^\mu\,,\,\gamma^\nu])_{ab}(\partial_\mu A_\nu - \partial_\nu A_\mu) + (\gamma^5)_{ab}d] \\ &\;\;\;\;\;+ \epsilon^{\mathcal{IJ}}{}_{\mathcal{K}}[i(\gamma^\mu)_{ab}(\partial_\mu A^{\mathcal{K}}) + (\gamma^5\gamma^\mu)_{ab}(\partial_\mu B^{\mathcal{K}}) - iC_{ab}F^{\mathcal{K}} - (\gamma^5)_{ab}G^{\mathcal{K}}]\label{eqn:vectorvaliserules2}
\end{align}

The resulting L and R matrices can be found in Appendix \ref{app:ExplicitLMatrices} in concurrence with \cite{calkins_is_2014}, and the corresponding Adinkra is shown below:

\begin{figure}[ht!]
    \centering
    \includegraphics[width = 0.6\textwidth]{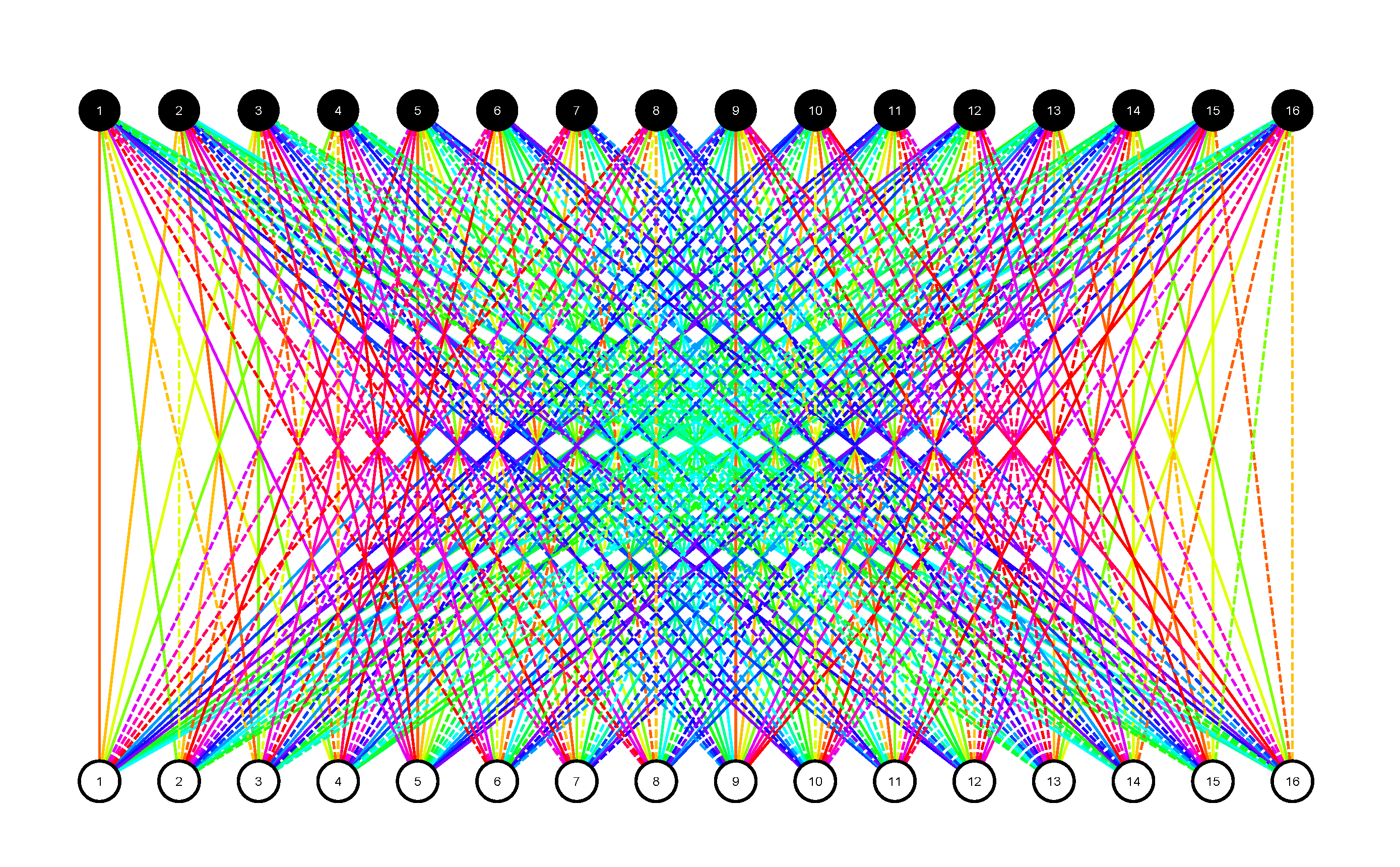}
    \caption{Adinkra for the 4D, $\mathcal{N}=4$ Maxwell supermultiplet with N=1 off shell closure: consists of 16 Boson (White) and 16 Fermion (Black) field components, and N $=16$ edge colors.}
    \label{fig:vectorvaliseadinkra}
\end{figure}

\subsection{Review of the 4D \texorpdfstring{$\mathcal{N}=4$}{TEXT} Vector-Tensor Supermultiplet}

$~~~~$ The 4D $\mathcal{N}=4$ Vector Tensor Multiplet with N=2 off shell closure is composed of a pair of 4D N=1 off shell chiral multiplets, a 4D N=1 off shell vector multiplet, and a 4D N=1 off shell tensor multiplet. This requires a pair of doublet operators $D^i_a,\;\Tilde{D}^i_a$. The transformation rules for this multiplet are given in \cite{gates_exploring_2019,gates_infinite-dimensional_2022} and are reproduced here. 

\begin{minipage}{0.49\textwidth}
\begin{align*}
    D_a^iA &= \delta^{ij}\Psi_a^j\;, \\
    D_a^iB &= i\delta^{ij}(\gamma^5)_a{}^b\Psi_b^j \;,\\
    D_a^iF &= \delta^{ij}(\gamma^\mu)_a{}^b\partial_\mu\Psi_b^j \;,\\
    D_a^iG &= i(\sigma^3)^{ij}(\gamma^5\gamma^\mu)_a{}^b\partial_\mu\Psi_b^j \;,\\
    D_a^iA_\mu &= i(\sigma^2)^{ij}(\gamma_\mu)_a{}^b\Psi_b^j \;,\\
    D_a^id &= i(\sigma^1)^{ij}(\gamma^5\gamma^\mu)_a{}^b\partial_\mu\Psi_b^j \;,\\
    D_a^i\tilde{A} &= (\sigma^3)^{ij}\tilde{\Psi}_a^j\;, \\
    D_a^i\tilde{B} &= i\delta^{ij}(\gamma^5)_a{}^b\tilde{\Psi}_b^j \;,\\
    D_a^i\tilde{F} &= \delta^{ij}(\gamma^\mu)_a{}^b\partial_\mu\tilde{\Psi}_b^j \;,\\
    D_a^i\tilde{G} &= i\delta^{ij}(\gamma^5\gamma^\mu)_a{}^b\partial_\mu\tilde{\Psi}_b^j \;,\\
    D_a^i\tilde{B}_{\mu\nu} &= -\frac{i}{4}(\sigma^2)^{ij}(\gamma_{[\mu}\gamma_{\nu]})_a{}^b\tilde{\Psi}_b^j \;,\\
    D_a^i\tilde{\phi} &= (\sigma^1)^{ij}\tilde{\Psi}_a^j \;,
\end{align*}
\end{minipage}\begin{minipage}{0.49\textwidth}
\begin{align}
	\tilde{D}_a^iA &= i(\sigma^2)^{ij}\tilde{\Psi}_a^j\;, \\
	\tilde{D}_a^iB &= -(\sigma^2)^{ij}(\gamma^5)_a{}^b\tilde{\Psi}_b^j \;,\\
	\tilde{D}_a^iF &= i(\sigma^2)^{ij}(\gamma^\mu)_a{}^b\partial_\mu\tilde{\Psi}_b^j \;,\\
	\tilde{D}_a^iG &= i(\sigma^1)^{ij}(\gamma^5\gamma^\mu)_a{}^b\partial_\mu\tilde{\Psi}_b^j \;,\\
	\tilde{D}_a^iA_\mu &= \delta^{ij}(\gamma_\mu)_a{}^b\tilde{\Psi}_b^j \;,\\
	\tilde{D}_a^id &= -i(\sigma^3)^{ij}(\gamma^5\gamma^\mu)_a{}^b\partial_\mu\tilde{\Psi}_b^j \;,\\
	\tilde{D}_a^i\tilde{A} &= -(\sigma^1)^{ij}\Psi_a^j\;, \\
	\tilde{D}_a^i\tilde{B} &= -(\sigma^2)^{ij}(\gamma^5)_a{}^b\Psi_b^j \;,\\
	\tilde{D}_a^i\tilde{F} &= i(\sigma^2)^{ij}(\gamma^\mu)_a{}^b\partial_\mu\Psi_b^j \;,\\
	\tilde{D}_a^i\tilde{G} &= -(\sigma^2)^{ij}(\gamma^5\gamma^\mu)_a{}^b\partial_\mu\Psi_b^j \;,\\
	\tilde{D}_a^i\tilde{B}_{\mu\nu} &= -\frac{1}{4}(\delta)^{ij}(\gamma_{[\mu}\gamma_{\nu]})_a{}^b\Psi_b^j \;,\\
	\tilde{D}_a^i\tilde{\phi} &= (\sigma^3)^{ij}\Psi_a^j \;,
\end{align}
\end{minipage}

\begin{align}
    D_a^i\Psi_b^j &= \delta^{ij}\{i(\gamma^\mu)_{ab}\partial_\mu A - (\gamma^5\gamma^\mu)_{ab}\partial_\mu B - iC_{ab}F\} \\
    &\; + (\gamma^5)_{ab}\{(\sigma^3)^{ij}G + (\sigma^1)^{ij}d\} + \frac14(\sigma^2)^{ij}(\gamma^{[\mu}\gamma^{\nu]})_{ab}F_{\mu\nu}\\
    D_a^i\tilde{\Psi}_b^j &= \delta^{ij}\{-(\gamma^5\gamma^\mu)_{ab}\partial_\mu \tilde{B} - iC_{ab}\tilde{F} + (\gamma^5)_{ab}\tilde{G}\} \\
    &\; +  i(\gamma^\mu)_{ab}\partial_\mu\{(\sigma^3)^{ij}\tilde{A} + (\sigma^1)^{ij}\tilde{\phi}\} - i(\sigma^2)^{ij}\epsilon_\mu{}^{\nu\alpha\beta}(\gamma^5\gamma^\mu)_{ab}\partial_\nu\tilde{B}_{\alpha\beta}\\
    \tilde{D}_a^i\Psi_b^j &= i(\sigma^2)^{ij}\{-(\gamma^5\gamma^\mu)_{ab}\partial_\mu \tilde{B} - iC_{ab}\tilde{F} + (\gamma^5)_{ab}\tilde{G}\} \\
    &\; +  i(\gamma^\mu)_{ab}\partial_\mu\{-(\sigma^1)^{ij}\tilde{A} + (\sigma^3)^{ij}\tilde{\phi}\} - \delta^{ij}\epsilon_\mu{}^{\nu\alpha\beta}(\gamma^5\gamma^\mu)_{ab}\partial_\nu\tilde{B}_{\alpha\beta}\\
    \tilde{D}_a^i\tilde{\Psi}_b^j &= i(\sigma^2)^{ij}\{i(\gamma^\mu)_{ab}\partial_\mu A - (\gamma^5\gamma^\mu)_{ab}\partial_\mu B - iC_{ab}F\} \\
    &\; + (\gamma^5)_{ab}\{(\sigma^1)^{ij}G + (\sigma^3)^{ij}d\} - \frac{i}{4}\delta^{ij}(\gamma^{[\mu}\gamma^{\nu]})_{ab}F_{\mu\nu}
\end{align}
\noindent
The L matrices for this theory are listed in Appendix \ref{app:ExplicitLMatrices}, and the Adinkra is shown below.

\begin{figure}[ht!]
    \centering
    \includegraphics[width = 0.6\textwidth]{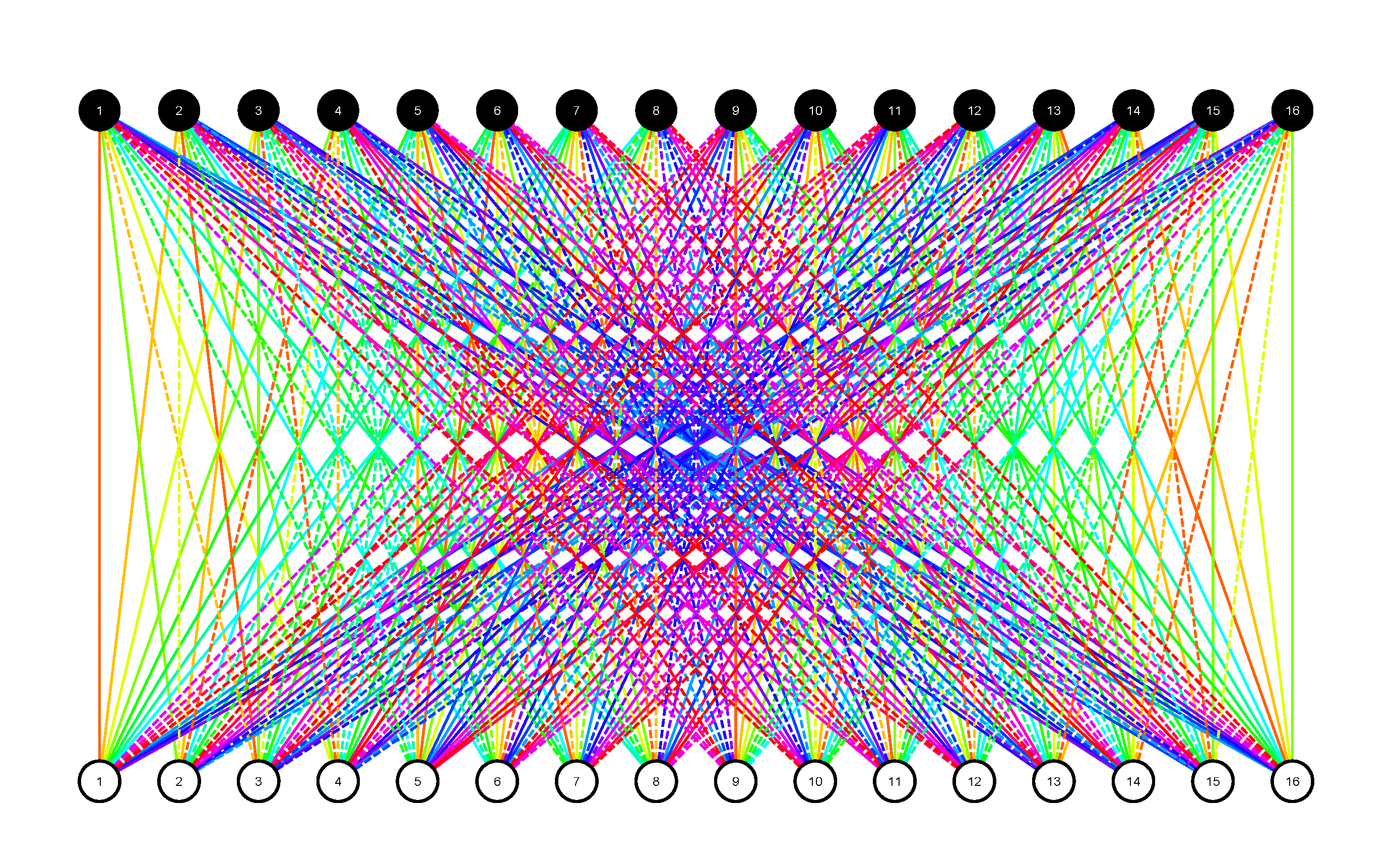}
    \caption{Adinkra for the 4D, $\mathcal{N}=4$ Vector-Tensor multiplet with $\mathcal{N}=2$ degrees of off-shell closure: consists of 16 Boson (White) and 16 Fermion (Black) field components, and N $=16$ edge colors.}
    \label{fig:tensorvaliseadinkra}
\end{figure}

It is also known that there exists a gauge duality between the 4D $\mathcal{N}=4$ Vector multiplet and the 4D $\mathcal{N}=4$ Vector-Tensor multiplet, and that the central charge structure is significantly simpler due to the changes in symmetry groups present. In this work, we report that the right-handed Hopping operators for both of these theories are the same, suggesting that a gauge duality in the 4D theories can be related to a left-permutation of the L-matrices along with a redefinition of sign factors. These Hopping operators are equal to the group $(\mathbb{S}_2)^4$.

\section{On-Shell Embedding}
\label{sec:embedding}

$~~~~$The off-shell problem has long remained a barrier between a complete understanding of SUSY theories, the resolution of which would have many implications \cite{gates_jr_fundamental_2001}. Recent work in polytopic representation has also been a formulation of attempts to simplify this problem, as were Garden algebras and codes. In the language of Adinkras, the off-shell problem would be posed as the following: given an on-shell valise Adinkra, does there exist an off-shell valise Adinkra in which the former is contained as a subgraph? Checking this for any two graphs, ie the subgraph Isomorphism Problem, is generally classified as NP-complete in complexity, but since Adinkras are a specific type of graph, there may be better ways to check for this algorithmically \cite{douglas_automorphism_2015}.

Hopping operators along with their connection to codes and the graph spectral methods used in Section \ref{sec:topologies}, are found to alleviate the complexity of the colored-graph isomorphism problem, due to their ability to classify the levels of equivalence between Adinkras. We now show how embedded topologies may present a route to approach these fundamental questions.

\subsection{Topology of the 4D \texorpdfstring{$\mathcal{N}=4$}{TEXT} Vector and Vector-Tensor multiplets}

$~~~~$In order to determine whether a theory can be embedded, it is natural to look to its topology first, since topologies are the most general fundamental structure of an Adinkra. Without utilizing a supercomputer or writing any especially crafty algorithm, we can find the topology of the two theories by performing the same operation as in Section \ref{sec:topologies}, only in a reverse manner. To motivate this, consider again the general form of the hopping operators \ref{eqn:Ncubehoppers}. Hopping operators of this form correspond to graphs with the same symmetries as a simple cubic lattice.  We are now considering two on-shell representations whose Adinkras are fully-connected bipartite graphs. For both $N=4$ and $N=8$, There are a single set of hopping operators of this form, which are the minimal off shell representations. The hopping operators associated with both theories are found to be isomorphic to the group $(\mathbb{Z}_2)^4$ itself, just as was the case for minimal $N=4$ and  $N=8$ multiplets.

To understand how two theories which differ by a quotient are related, we draw from the original works once more \cite{doran_topology_2008}. There are several ways to describe the quotient: It can be thought of as the assigning of cyclical orbits between pairs of vertices, those which will be self-identified in the resulting graph, along with the edges that connect them. The code can also be thought of as a set of vectors which are used to perform reflections about their midpoints. Illustrations in the referenced work suggest this can be thought of as an orthographic projection of one half of the graph onto the inverse of its other half. It is a process of folding and reassigning which \textit{breaks} the topology of the original graph to create a more compact topology, or ``eliminate redundancy". To aid in the following discussion, we have reproduced those figures below.

\begin{figure}[ht!]
    \centering
    \includegraphics[width=0.35\linewidth]{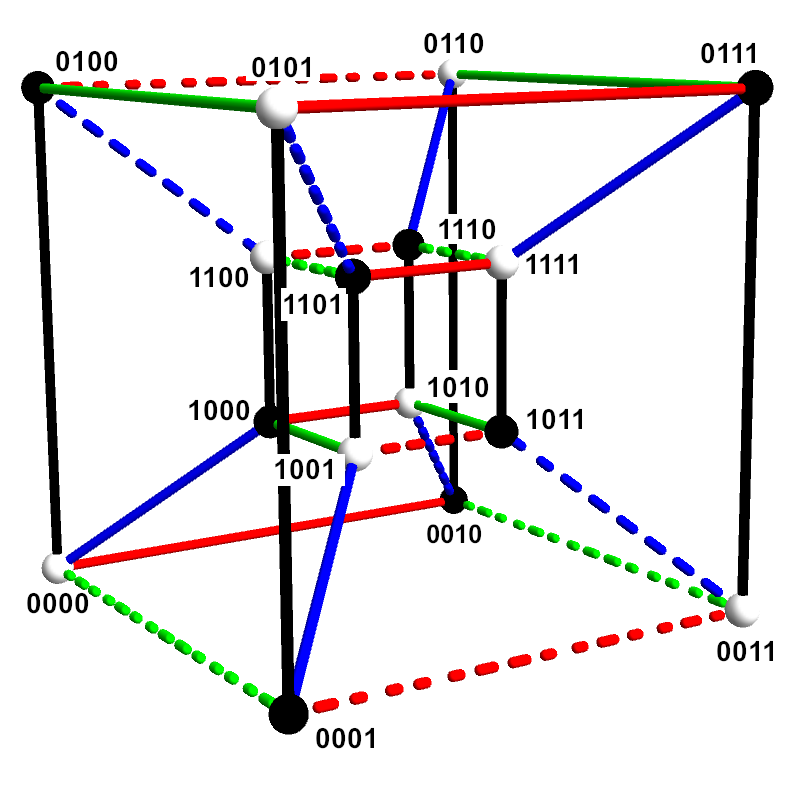}\hspace{0.5cm}
    \raisebox{3cm}{$\rightarrow$}\hspace{0.5cm}
    \raisebox{0.5cm}{\includegraphics[width=0.3\linewidth]{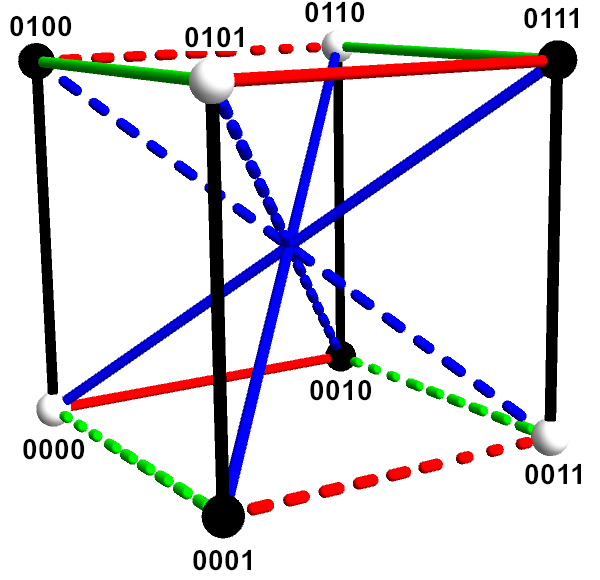}}
    \caption{The quotient operation on a 4-cube representation to derive a minimal N=4 representation. The code used in this quotient is the only doubly even code of length 4: (1111). Every vertex is identified with the vertex found by direct sum of this code modulo 2. The result is a fully-connected bipartite graph, which is the most compact topology allowed in this case. Further details can be found in \cite{doran_topology_2008}.}
    \label{fig:4-Cube-quotient }
\end{figure}

It is important to note that these two objects are strongly related: in the first image, one can see that by inverting the inner cube and enlarging it so that it overlaps with the outer cube, we recover the second. The nature of this operation is the same for every iteration it is performed. In the image above, no information about edge color or dashing is lost in the quotient, because with respect to the doubly even code, this information is twice redundant and is hence preserved during the quotient. It is determined in \cite{doran_relating_2008} that this is generally the case for quotients of N-cubes by doubly even codes. What can we say about the former containing the latter as a subgraph? Although the quotiented graph is a partition graph of the former--and therefore it can be said that all information in the quotiented graph exists in the original, there is no direct embedding of these distinct topologies so it cannot be stated that one is a subgraph of the other. The act of introducing new nodes would not be enough to recover the original from the quotient; one instead has to devise an encoding of the theory using the doubly even code; and although this may be possible, it is distinct from traditionally sought-after embedding, so it is not explored further in this work. 

We now show the connection between the topology of the Maxwell supermultiplet and that of off-shell theories like the Spinning Particle multiplet. It is found that by the application of the inverse quotient methods, we can directly arrive at the code used to generate the theory. The method is as follows: First, derive the Hopping operators of the representation, and assuming that they are in a form given by \ref{eqn:Ncubehoppers}, list them in the form of their vectors $\bm{a}^{(\text{I})}$. From this, read the representation as a code $C^{\mathcal{R}}$ of length N and dimension k, where the matrix dimension $d = 2^k$. There now exists an even code $C$ which is found in the same way, as a quotient of the N-cube by $C^{\mathcal{R}}$ and a reading of its vector notation.  

The same end is accomplished by quotienting the N-cube further than discussed in Section \ref{ssec:decoding}, using a new code which is not doubly even: the well known Reed-Muller code $\text{RM}(2,4)$. Given the dimensions of this code, it is apparent that the encoded information necessary to convey off-shell supersymmetry becomes lost following its quotient. To be precise, the ``rate" of information loss of a code is the ratio of its dimension to its length: $\frac{k}{n}$. This code then has a rate of $\frac{11}{16}$, while the maximum rate for preserving off-shell closure was observed to be $50\%$ in \cite{doran_relating_2008}. The generator of this code is given below:
\begin{align}
    \text{RM}(2,4):\;\; \begin{tabular}{c}
        1111111111111111\\
        0000000011111111\\
        0000111100001111\\
        0011001100110011\\
        0101010101010101\\
        0000000000001111\\
        0000000000110011\\
        0000000001010101\\
        0000001100000011\\
        0000010100000101\\
        0001000100010001
    \end{tabular}
\end{align}

The code above performs $2^{11}$ quotients on the 16-cube, leaving a total of 32 nodes. The result of this quotient is the chromotopology of a minimal on shell valise representation. The resulting representation has hopping operators which are isomorphic to $(\mathbb{Z}_2)^4$, exactly as the Vector and Vector-Tensor multiplets. The implication is that the Hopping operators of the Maxwell supermultiplet can be derived from any off-shell N $=16$ representation, because all off shell representations are derived from subcodes of this code, and thus contain redundancies of this supermultiplet. It can therefore be concluded that the on-shell Adinkra topology considered herein cannot be found as a subgraph in that of an off-shell Adinkra, because the topologies \textit{can} be related through the act of breaking and folding. If in principle there were ways to preserve the information lost upon performing a quotient, then there may be a different way in which we can approach this question \cite{gates_jr_superfield_2020,gates_advening_2020,gates_component_2021}.

We now turn our focus to the 10D and 4D abelian Vector supermultiplets. By ruling out the embedding of the Maxwell supermultiplet in a higher dimensional off-shell theory, we still have not proven any such ruling for these two theories. It would be natural to continue looking for solutions with the methods discussed here, but we must develop a more thoughtful reasoning first. We know that minimal N $=16$ off-shell theories come in Pauli matrix representations such as the Spinning particle multiplet, and also that obtaining one of these representations by augmentation of the $9\times16$ L-matrices would involve finding an instance of these matrices embedded. 

We may first want to ask, how many new representations can be formed from the abelian Vector supermultiplet's L-matrices which are topologically distinct from that of the Maxwell supermultiplet? 

\section{Conclusions}
\label{sec:conclusion}

$~~~~$In this work we have made large strides towards a more complete understanding of N=16 Genomics. We first find that contrary to expectations, the magic numbers of the derived hopping operators and the spinning particle multiplet are all different, making the polytopic representation of such theories more compelling and open. Expanding on the rich and insightful bodies of work \cite{doran_codes_2011,doran_relating_2008,doran_topology_2008,gates_n4_2016}, we implement the ideas of quotienting by codes in a computational-experimental effort towards tackling the off-shell problem for 4D $\mathcal{N}=4$ theories. We show that by using codes and quotients of the N-cube, we are able to derive and classify possible topologies for off-shell and on-shell representations.

We find by a comparison of graph spectra that the 1D N=16 Spinning Particle multiplet has the topology $E_8\times E_8$, and derive the hopping operators which follow directly from the two possible topologies, $E_8\times E_8$ and $D_{16}$. We have then revisited several on-shell theories and derive their L-matrices and hopping operators, the latter which are found to be equivalent. We report that these hopping operators can be derived using a similar approach to the one stated prior, thus suggesting that the off shell problem is decidedly closed for the case of the 4D $\mathcal{N}=4$ Maxwell supermultiplet.

Thus, there is only one remaining avenue that suggests itself for further investigation.

As we noted in the introduction, the 4D, $\cal N$ = 4 superfiber bundle approach can be taken as starting  point.  In particular, after imposing the constraint that defines the vectorial connection in terms of the spinorial one.  Since the latter of these is by definition a superfield, it must contain all of the component fields necessary
to describe an off-shell.  So the next goal must be apply the technology of adinkras developed in the works of 
[\citen{gates_jr_superfield_2020}, \citen{gates_advening_2020}, \citen{gates_component_2021}].  This should lead to a description
of the 4D, $\cal N$ = 4 gauge multiplet that is the equivalent to the 4D, $\cal N$ = 1 supergravity supermultiplet, {\it {with no reference to an action}}.  As with
the Breitenlohner \cite{BREITENLOHNER197749} description, additional study would be required to breach this final obstacle.

\subsection*{Acknowledgements}
$~~~~$The research was supported during a portion of the time it was carried out by the endowment of the Ford Foundation Professorship of Physics at Brown 
University. Jude Bedessem, Jeth Arunseangroj, Gabriel Yerger, Devin Bristow, and Shane Weiner would like to acknowledge their
participation in the SSTPRS program. As well,
we thank the latter two for participation in
this project's early stages, and Kevin Iga for multiple illuminating discussions on adinkras and codes.
All also gratefully acknowledge the support of the Brown Theoretical Physics Center. 

In addition, the research of S.J.G. wishes to acknowledge the endowment of the Ford Foundation Professor of Physics, and appointments as an 
Affiliate Professor of Mathematics, Department of Physics, and 
Faculty Fellow, Watson Institute for International Studies \& Public Affairs,
Brown University, July 2017-June 30, 2022
is currently supported by the Clark Leadership Chair in Science endowment at the University of Maryland - College Park.

\newpage

\appendix

\section{Explicit L-Matrices}
\label{app:ExplicitLMatrices}

Explicit L-matrices are given for each of the theories which have square matrices, ie. equal numbers of bosons and fermions. For the theories which do not have square matrices, we leave out explicit representations, but these can be pulled from the \href{https://github.com/Gwyerger/SUSYHelper}{SUSYHelper package}.

\subsection{4D \texorpdfstring{$\mathcal{N}=4$}{TEXT} Maxwell Supermultiplet}
\renewcommand{\arraystretch}{1.1}
The L Matrices of the 1D N=16 Maxwell Supermultiplet with $N=1$ off-shell closure are:
\begin{align*}
    L_1 &= \left[ \;\;\begin{NiceArray}{w{c}{\mylen}w{c}{\mylen}w{c}{\mylen}w{c}{\mylen}}
    (10)_b(243) & 0 & 0 & 0  \\
    0 & (10)_b(243) & 0 & 0 \\
    0 & 0 & (10)_b(243) & 0 \\
    0 & 0 & 0 & (10)_b(1243)
\end{NiceArray} \;\;\right] &
L_2 &= \left[ \;\;\begin{NiceArray}{w{c}{\mylen}w{c}{\mylen}w{c}{\mylen}w{c}{\mylen}}
    (12)_b(123) & 0 & 0 & 0  \\
    0 & (12)_b(123) & 0 & 0 \\
    0 & 0 & (12)_b(123) & 0 \\
    0 & 0 & 0 & (12)_b(23)
\end{NiceArray}\;\;\right]\\
L_3 &= \left[ \;\;\begin{NiceArray}{w{c}{\mylen}w{c}{\mylen}w{c}{\mylen}w{c}{\mylen}}
    (6)_b(134) & 0 & 0 & 0  \\
    0 & (6)_b(134) & 0 & 0 \\
    0 & 0 & (6)_b(134) & 0 \\
    0 & 0 & 0 & (0)_b(14)
\end{NiceArray}\;\;\right] &
L_4 &= \left[ \;\;\begin{NiceArray}{w{c}{\mylen}w{c}{\mylen}w{c}{\mylen}w{c}{\mylen}}
    (0)_b(142) & 0 & 0 & 0  \\
    0 & (0)_b(142) & 0 & 0 \\
    0 & 0 & (0)_b(142) & 0 \\
    0 & 0 & 0 & (6)_b(1342)
\end{NiceArray}\;\;\right]\\
L_{5} &= \left[ \;\;\begin{NiceArray}{w{c}{\mylen}w{c}{\mylen}w{c}{\mylen}w{c}{\mylen}}
    0 & (15)_b(243) & 0 & 0  \\
    (0)_b(243) & 0 & 0 & 0 \\
    0 & 0 & 0 & (2)_b(243) \\
    0 & 0 & (13)_b(1243) & 0
\end{NiceArray}\;\;\right] &
L_{6} &= \left[ \;\;\begin{NiceArray}{w{c}{\mylen}w{c}{\mylen}w{c}{\mylen}w{c}{\mylen}}
    0 & (9)_b(123) & 0 & 0  \\
    (6)_b(123) & 0 & 0 & 0 \\
    0 & 0 & 0 & (4)_b(123) \\
    0 & 0 &(11)_b(23) & 0
\end{NiceArray}\;\;\right]\\
L_{7} &= \left[ \;\;\begin{NiceArray}{w{c}{\mylen}w{c}{\mylen}w{c}{\mylen}w{c}{\mylen}}
    0 & (3)_b(134) & 0 & 0  \\
    (12)_b(134) & 0 & 0 & 0 \\
    0 & 0 & 0 & (14)_b(134) \\
    0 & 0 & (1)_b(1342) & 0
\end{NiceArray}\;\;\right] &
L_{8} &= \left[ \;\;\begin{NiceArray}{w{c}{\mylen}w{c}{\mylen}w{c}{\mylen}w{c}{\mylen}}
    0 & (5)_b(142) & 0 & 0  \\
    (10)_b(142) & 0 & 0 & 0 \\
    0 & 0 & 0 & (8)_b(142) \\
    0 & 0 & (1)_b(1342) & 0
\end{NiceArray}\;\;\right]\\
L_9 &= \left[ \;\;\begin{NiceArray}{w{c}{\mylen}w{c}{\mylen}w{c}{\mylen}w{c}{\mylen}}
    0 & 0 & (0)_b(243) & 0  \\
    0 & 0 & 0 & (2)_b(243) \\
    (15)_b(243) & 0 & 0 & 0 \\
    0 & (13)_b(1243) & 0 & 0
\end{NiceArray}\;\;\right] &
L_{10} &= \left[ \;\;\begin{NiceArray}{w{c}{\mylen}w{c}{\mylen}w{c}{\mylen}w{c}{\mylen}}
    0 & 0 & (6)_b(123) & 0  \\
    0 & 0 & 0 & (4)_b(123) \\
    (9)_b(123) & 0 & 0 & 0 \\
    0 & (11)_b(23) & 0 & 0
\end{NiceArray}\;\;\right]\\
L_{11} &= \left[ \;\;\begin{NiceArray}{w{c}{\mylen}w{c}{\mylen}w{c}{\mylen}w{c}{\mylen}}
    0 & 0 & (12)_b(134) & 0  \\
    0 & 0 & 0 & (14)_b(134) \\
    (3)_b(134) & 0 & 0 & 0 \\
    0 & (7)_b(14) & 0 & 0
\end{NiceArray}\;\;\right] &
L_{12} &= \left[ \;\;\begin{NiceArray}{w{c}{\mylen}w{c}{\mylen}w{c}{\mylen}w{c}{\mylen}}
    0 & 0 & (10)_b(142) & 0  \\
    0 & 0 & 0 & (8)_b(142) \\
    (5)_b(142) & 0 & 0 & 0 \\
    0 & (1)_b(1342) & 0 & 0
\end{NiceArray}\;\;\right]
\end{align*}

\newpage

\begin{align*}
L_{13} &= \left[ \;\;\begin{NiceArray}{w{c}{\mylen}w{c}{\mylen}w{c}{\mylen}w{c}{\mylen}}
    0 & 0 & 0 & (2)_b(243)  \\
    0 & 0 & (15)_b(243) & 0 \\
    0 & (0)_b(243) & 0 & 0 \\
    (13)_b(1243) & 0 & 0 & 0
\end{NiceArray}\;\;\right] &
L_{14} &= \left[ \;\;\begin{NiceArray}{w{c}{\mylen}w{c}{\mylen}w{c}{\mylen}w{c}{\mylen}}
    0 & 0 & 0 & (4)_b(123)  \\
    0 & 0 & (9)_b(123) & 0 \\
    0 & (6)_b(123) & 0 & 0 \\
    (11)_b(23) & 0 & 0 & 0
\end{NiceArray}\;\;\right]\\
L_{15} &= \left[ \;\;\begin{NiceArray}{w{c}{\mylen}w{c}{\mylen}w{c}{\mylen}w{c}{\mylen}}
    0 & 0 & 0 & (14)_b(134)  \\
    0 & 0 & (3)_b(134) & 0 \\
    0 & (12)_b(134) & 0 & 0 \\
    (7)_b(14) & 0 & 0 & 0
\end{NiceArray}\;\;\right] &
L_{16} &= \left[ \;\;\begin{NiceArray}{w{c}{\mylen}w{c}{\mylen}w{c}{\mylen}w{c}{\mylen}}
    0 & 0 & 0 & (8)_b(142)  \\
    0 & 0 & (5)_b(142) & 0 \\
    0 & (10)_b(142) & 0 & 0 \\
    (1)_b(1342) & 0 & 0 & 0
\end{NiceArray}\;\;\right]
\end{align*}

\subsection{4D \texorpdfstring{$\mathcal{N}=4$}{TEXT} Vector-Tensor Supermultiplet}

The L-matrices for the 1D N=16 Vector-Tensor Supermultiplet with N=2 off-shell closure are:

\begin{align*}
    L_1 &= \left[ \;\;\begin{NiceArray}{w{c}{\mylen}w{c}{\mylen}w{c}{\mylen}w{c}{\mylen}}
    (10)_b(243) & 0 & 0 & 0  \\
    0 & (10)_b(1243) & 0 & 0 \\
    0 & 0 & (10)_b(243) & 0 \\
    0 & 0 & 0 & (7)_b(1324)
\end{NiceArray}\;\;\right] &
L_2 &= \left[ \;\;\begin{NiceArray}{w{c}{\mylen}w{c}{\mylen}w{c}{\mylen}w{c}{\mylen}}
    (12)_b(123) & 0 & 0 & 0  \\
    0 & (12)_b(23) & 0 & 0 \\
    0 & 0 & (12)_b(123) & 0 \\
    0 & 0 & 0 & (2)_b(1423)
\end{NiceArray}\;\;\right]\\
L_3 &= \left[ \;\;\begin{NiceArray}{w{c}{\mylen}w{c}{\mylen}w{c}{\mylen}w{c}{\mylen}}
    (6)_b(134) & 0 & 0 & 0  \\
    0 & (0)_b(14) & 0 & 0 \\
    0 & 0 & (6)_b(134) & 0 \\
    0 & 0 & 0 & (4)_b(34)
\end{NiceArray}\;\;\right] &
L_4 &= \left[ \;\;\begin{NiceArray}{w{c}{\mylen}w{c}{\mylen}w{c}{\mylen}w{c}{\mylen}}
    (0)_b(142) & 0 & 0 & 0  \\
    0 & (6)_b(1342) & 0 & 0 \\
    0 & 0 & (0)_b(142) & 0 \\
    0 & 0 & 0 & (1)_b(12)
\end{NiceArray}\;\;\right]\\
L_{5} &= \left[ \;\;\begin{NiceArray}{w{c}{\mylen}w{c}{\mylen}w{c}{\mylen}w{c}{\mylen}}
    0 & (2)_b(243) & 0 & 0  \\
    (13)_b(1243) & 0 & 0 & 0 \\
    0 & 0 & 0 & (11)_b(243) \\
    0 & 0 & (0)_b(1324) & 0
\end{NiceArray}\;\;\right] &
L_{6} &= \left[ \;\;\begin{NiceArray}{w{c}{\mylen}w{c}{\mylen}w{c}{\mylen}w{c}{\mylen}}
    0 & (4)_b(123) & 0 & 0  \\
    (11)_b(23) & 0 & 0 & 0 \\
    0 & 0 & 0 & (13)_b(123) \\
    0 & 0 &(5)_b(1423) & 0
\end{NiceArray}\;\;\right]\\
L_{7} &= \left[ \;\;\begin{NiceArray}{w{c}{\mylen}w{c}{\mylen}w{c}{\mylen}w{c}{\mylen}}
    0 & (14)_b(134) & 0 & 0  \\
    (7)_b(14) & 0 & 0 & 0 \\
    0 & 0 & 0 & (7)_b(134) \\
    0 & 0 & (3)_b(34) & 0
\end{NiceArray}\;\;\right] &
L_{8} &= \left[ \;\;\begin{NiceArray}{w{c}{\mylen}w{c}{\mylen}w{c}{\mylen}w{c}{\mylen}}
    0 & (8)_b(142) & 0 & 0  \\
    (1)_b(1342) & 0 & 0 & 0 \\
    0 & 0 & 0 & (1)_b(142) \\
    0 & 0 & (6)_b(12) & 0
\end{NiceArray}\;\;\right]
\end{align*}

\begin{align*}
L_{9} &= \left[ \;\;\begin{NiceArray}{w{c}{\mylen}w{c}{\mylen}w{c}{\mylen}w{c}{\mylen}}
    0 & 0 & (13)_b(243) & 0  \\
    0 & 0 & 0 & (10)_b(1243) \\
    (5)_b(243) & 0 & 0 & 0 \\
    0 & (15)_b(1324) & 0 & 0
\end{NiceArray}\;\;\right] &
L_{10} &= \left[ \;\;\begin{NiceArray}{w{c}{\mylen}w{c}{\mylen}w{c}{\mylen}w{c}{\mylen}}
    0 & 0 & (11)_b(123) & 0  \\
    0 & 0 & 0 & (12)_b(23) \\
    (3)_b(123) & 0 & 0 & 0 \\
    0 & (10)_b(1423) & 0 & 0
\end{NiceArray}\;\;\right]\\
L_{11} &= \left[ \;\;\begin{NiceArray}{w{c}{\mylen}w{c}{\mylen}w{c}{\mylen}w{c}{\mylen}}
    0 & 0 & (1)_b(134) & 0  \\
    0 & 0 & 0 & (0)_b(14) \\
    (9)_b(134) & 0 & 0 & 0 \\
    0 & (12)_b(34) & 0 & 0
\end{NiceArray}\;\;\right] &
L_{12} &= \left[ \;\;\begin{NiceArray}{w{c}{\mylen}w{c}{\mylen}w{c}{\mylen}w{c}{\mylen}}
    0 & 0 & (7)_b(142) & 0  \\
    0 & 0 & 0 & (6)_b(1342) \\
    (15)_b(142) & 0 & 0 & 0 \\
    0 & (9)_b(12) & 0 & 0
\end{NiceArray}\;\;\right]\\
L_{13} &= \left[ \;\;\begin{NiceArray}{w{c}{\mylen}w{c}{\mylen}w{c}{\mylen}w{c}{\mylen}}
    0 & 0 & 0 & (10)_b(243)  \\
    0 & 0 & (2)_b(1243) & 0 \\
    0 & (11)_b(243) & 0 & 0 \\
    (7)_b(1324) & 0 & 0 & 0
\end{NiceArray}\;\;\right] &
L_{14} &= \left[ \;\;\begin{NiceArray}{w{c}{\mylen}w{c}{\mylen}w{c}{\mylen}w{c}{\mylen}}
    0 & 0 & 0 & (12)_b(123)  \\
    0 & 0 & (4)_b(23) & 0 \\
    0 & (13)_b(123) & 0 & 0 \\
    (2)_b(1423) & 0 & 0 & 0
\end{NiceArray}\;\;\right]\\
L_{15} &= \left[ \;\;\begin{NiceArray}{w{c}{\mylen}w{c}{\mylen}w{c}{\mylen}w{c}{\mylen}}
    0 & 0 & 0 & (6)_b(134)  \\
    0 & 0 & (8)_b(14) & 0 \\
    0 & (7)_b(134) & 0 & 0 \\
    (4)_b(34) & 0 & 0 & 0
\end{NiceArray}\;\;\right] &
L_{16} &= \left[ \;\;\begin{NiceArray}{w{c}{\mylen}w{c}{\mylen}w{c}{\mylen}w{c}{\mylen}}
    0 & 0 & 0 & (0)_b(142)  \\
    0 & 0 & (14)_b(1342) & 0 \\
    0 & (1)_b(142) & 0 & 0 \\
    (1)_b(12) & 0 & 0 & 0
\end{NiceArray}\;\;\right]
\end{align*}

\newpage
\subsection{1D, \texorpdfstring{$\mathcal{N}=16$}{TEXT} Spinning Particle Multiplet}

The following matrices were calculated using the recursive formula found in \cite{gates_theory_1996} and by applying the convention $\textbf{L}_I - \textbf{R}_I^{\text{T}} = 0$
\begin{align*}
    &\textbf{L}_{1}&&= && \text{I} && \otimes && \text{I} && \otimes && \text{I} && \otimes && \text{I} && \otimes && \text{I} && \otimes && \text{I} && \otimes && \text{I} && = && (\textbf{R}_{1})^{\text{T}} \quad ;\\
    &\textbf{L}_2&&=&& i\sigma^3 && \otimes && \text{I} && \otimes && \sigma^3 && \otimes && \sigma^2 && \otimes && \text{I} && \otimes && \text{I} && \otimes && \text{I} && = && (\textbf{R}_{2})^{\text{T}} \quad ;\\
    &\textbf{L}_3&&=&& i\sigma^3 && \otimes && \sigma^3 && \otimes && \sigma^2 && \otimes && \text{I} && \otimes && \text{I} && \otimes && \text{I} && \otimes && \text{I} && = && (\textbf{R}_3)^{\text{T}} \quad ;\\
    &\textbf{L}_4&&=&& i\sigma^3 && \otimes && \text{I} && \otimes && \sigma^1 && \otimes && \sigma^2 && \otimes && \text{I} && \otimes && \text{I} && \otimes && \text{I} && = && (\textbf{R}_4)^{\text{T}} \quad ;\\
    &\textbf{L}_5&&=&& i\sigma^3 && \otimes && \sigma^2 && \otimes && \text{I} && \otimes && \sigma^3 && \otimes && \text{I} && \otimes && \text{I} && \otimes && \text{I} && = && (\textbf{R}_5)^{\text{T}} \quad ;\\
    &\textbf{L}_6&&=&& i\sigma^3 && \otimes && \sigma^2 && \otimes && \text{I} && \otimes && \sigma^1 && \otimes && \text{I} && \otimes && \text{I} && \otimes && \text{I} && = && (\textbf{R}_6)^{\text{T}} \quad ;\\
    &\textbf{L}_7&&=&& i\sigma^3 && \otimes && \sigma^1 && \otimes && \sigma^2 && \otimes && \text{I} && \otimes && \text{I} && \otimes && \text{I} && \otimes && \text{I} && = && (\textbf{R}_7)^{\text{T}} \quad ;\\
    &\textbf{L}_8&&=&& i\sigma^3 && \otimes && \sigma^2 && \otimes && \sigma^2 && \otimes && \sigma^2 && \otimes && \text{I} && \otimes && \text{I} && \otimes && \text{I} && = && (\textbf{R}_8)^{\text{T}} \quad ;\\
    &\textbf{L}_9&&=&& i\sigma^2 && \otimes && \text{I} && \otimes && \text{I} && \otimes && \text{I} && \otimes && \text{I} && \otimes && \text{I} && \otimes && \text{I} && = && (\textbf{R}_9)^{\text{T}} \quad ;\\
    &\textbf{L}_{10}&&=&&i\sigma^1 && \otimes && \text{I} && \otimes && \text{I} && \otimes && \text{I} && \otimes && \text{I} && \otimes && \sigma^3 && \otimes && \sigma^2 && = && (\textbf{R}_{10})^{\text{T}} \quad ;\\
    &\textbf{L}_{11}&&=&&i\sigma^1 && \otimes && \text{I} && \otimes && \text{I} && \otimes && \text{I} && \otimes && \sigma^3 && \otimes && \sigma^2 && \otimes && \text{I} && = && (\textbf{R}_{11})^{\text{T}} \quad ;\\
    &\textbf{L}_{12}&&=&& i\sigma^1 && \otimes && \text{I} && \otimes && \text{I} && \otimes && \text{I} && \otimes && \text{I} && \otimes && \sigma^1 && \otimes && \sigma^2 && = && (\textbf{R}_{12})^{\text{T}} \quad
    \end{align*}
    \begin{align*}
    &\textbf{L}_{13}&&=&& i\sigma^1 && \otimes && \text{I} && \otimes && \text{I} && \otimes && \text{I} && \otimes && \sigma^2 && \otimes && \text{I} && \otimes && \sigma^3 && = && (\textbf{R}_{13})^{\text{T}} \quad ;\\
    &\textbf{L}_{14}&&=&& i\sigma^1 && \otimes && \text{I} && \otimes && \text{I} && \otimes && \text{I} && \otimes && \sigma^2 && \otimes && \text{I} && \otimes && \sigma^1 && = && (\textbf{R}_{14})^{\text{T}} \quad ;\\
    &\textbf{L}_{15}&&=&& i\sigma^1 && \otimes && \text{I} && \otimes && \text{I} && \otimes && \text{I} && \otimes && \sigma^1 && \otimes && \sigma^2 && \otimes && \text{I} && = && (\textbf{R}_{15})^{\text{T}} \quad ;\\
    &\textbf{L}_{16}&&=&& i\sigma^1 && \otimes && \text{I} && \otimes && \text{I} && \otimes && \text{I} && \otimes && \sigma^2 && \otimes && \sigma^2 && \otimes && \sigma^2 && = && (\textbf{R}_{16})^{\text{T}} \quad ;
\end{align*}

\section{Explicit Hopping Operators}
\label{app:ExplicitHoppers}

Representations of explicit hopping operators for the theories with square L-matrices (valise) are given here. By convention \cite{cianciara_cal_2023}, we have displayed the right-handed hopping operators.

\subsection{4D \texorpdfstring{$\mathcal{N}=4$}{TEXT} On-Shell Multiplets}

The hopping operators for both the 4D $\mathcal{N}=4$ vector multiplet and the 4D $\mathcal{N}=4$ vector-tensor multiplet are the same up to a reordering. These are:

\begin{minipage}{0.49\textwidth}
\begin{align*}
&H_1& &=& &\text{I}& &\otimes& &\text{I}& &\otimes& &\text{I}& &\otimes& &\text{I} \quad;\\ 
    &H_2& &=& &\text{I}& &\otimes& &\text{I}& &\otimes& &\text{I}& &\otimes& &\sigma^1 \quad;\\ 
    &H_3& &=& &\text{I}& &\otimes& &\text{I}& &\otimes& &\sigma^1& &\otimes& &\text{I} \quad;\\ 
    &H_4& &=& &\text{I}& &\otimes& &\text{I}& &\otimes& &\sigma^1& &\otimes& &\sigma^1 \quad;\\ 
    &H_5& &=& &\text{I}& &\otimes& &\sigma^1& &\otimes& &\text{I}& &\otimes& &\text{I} \quad;\\ 
    &H_6& &=& &\text{I}& &\otimes& &\sigma^1& &\otimes& &\text{I}& &\otimes& &\sigma^1 \quad;\\ 
    &H_7& &=& &\text{I}& &\otimes& &\sigma^1& &\otimes& &\sigma^1& &\otimes& &\text{I} \quad;\\ 
    &H_8& &=& &\text{I}& &\otimes& &\sigma^1& &\otimes& &\sigma^1& &\otimes& &\sigma^1 \quad;
\end{align*}
\end{minipage}
\begin{minipage}{0.49\textwidth}
\begin{align*}
    &H_{9}& &=& &\sigma^1& &\otimes& &\text{I}& &\otimes& &\text{I}& &\otimes& &\text{I} \quad;\\ 
    &H_{10}& &=& &\sigma^1& &\otimes& &\text{I}& &\otimes& &\text{I}& &\otimes& &\sigma^1 \quad;\\ 
    &H_{11}& &=& &\sigma^1& &\otimes& &\text{I}& &\otimes& &\sigma^1& &\otimes& &\text{I} \quad;\\ 
    &H_{12}& &=& &\sigma^1& &\otimes& &\text{I}& &\otimes& &\sigma^1& &\otimes& &\sigma^1 \quad;\\
    &H_{13}& &=& &\sigma^1& &\otimes& &\sigma^1& &\otimes& &\text{I}& &\otimes& &\text{I} \quad;\\ 
    &H_{14}& &=& &\sigma^1& &\otimes& &\sigma^1& &\otimes& &\text{I}& &\otimes& &\sigma^1 \quad;\\ 
    &H_{15}& &=& &\sigma^1& &\otimes& &\sigma^1& &\otimes& &\sigma^1& &\otimes& &\text{I} \quad;\\ 
    &H_{16}& &=& &\sigma^1& &\otimes& &\sigma^1& &\otimes& &\sigma^1& &\otimes& &\sigma^1 \quad; 
\end{align*}
\end{minipage}

\subsection{1D \texorpdfstring{$\mathcal{N}=16$}{TEXT} Spinning Particle Multiplet}

\begin{align*}
    &H_{1}&&= && \text{I} && \otimes && \text{I} && \otimes && \text{I} && \otimes && \text{I} && \otimes && \text{I} && \otimes && \text{I} && \otimes && \text{I} &\quad; \\
    &H_2&&= &&\text{I} && \otimes && \text{I} && \otimes && \text{I} && \otimes && \sigma^1 && \otimes && \text{I} && \otimes && \text{I} && \otimes && \text{I} &\quad; \\
    &H_3& &= && \text{I} && \otimes && \text{I} && \otimes && \sigma^1 && \otimes && \text{I} && \otimes && \text{I} && \otimes && \text{I} && \otimes && \text{I} &\quad; \\
    &H_4&&=&& \text{I} && \otimes && \text{I} && \otimes && \sigma^1 && \otimes && \sigma^1 && \otimes && \text{I} && \otimes && \text{I} && \otimes && \text{I} &\quad; \\
    &H_5&&=&& \text{I} && \otimes && \sigma^1 && \otimes && \text{I} && \otimes && \text{I} && \otimes && \text{I} && \otimes && \text{I} && \otimes && \text{I}  &\quad; \\
    &H_6&&= && \text{I} && \otimes && \sigma^1 && \otimes && \text{I} && \otimes && \sigma^1 && \otimes && \text{I} && \otimes && \text{I} && \otimes && \text{I} &\quad; \\
    &H_7&&=&& \text{I} && \otimes && \sigma^1 && \otimes && \sigma^1 && \otimes && \text{I} && \otimes && \text{I} && \otimes && \text{I} && \otimes && \text{I} &\quad; \\
    &H_8&&=&& \text{I} && \otimes && \sigma^1 && \otimes && \sigma^1 && \otimes && \sigma^1 && \otimes && \text{I} && \otimes && \text{I} && \otimes && \text{I}  &\quad
    \end{align*}
    \begin{align*}
    &H_9&&=&& \sigma^1 && \otimes && \text{I} && \otimes && \text{I} && \otimes && \text{I} && \otimes && \text{I} && \otimes && \text{I} && \otimes && \text{I}  &\quad; \\
    &H_{10}&&=&& \sigma^1 && \otimes && \text{I} && \otimes && \text{I} && \otimes && \text{I} && \otimes && \text{I} && \otimes && \text{I} && \otimes && \sigma^1  &\quad; \\
    &H_{11}&&= && \sigma^1 && \otimes && \text{I} && \otimes && \text{I} && \otimes && \text{I} && \otimes && \text{I} && \otimes && \sigma^1 && \otimes && \text{I} &\quad; \\
    &H_{12}&&= && \sigma^1 && \otimes && \text{I} && \otimes && \text{I} && \otimes && \text{I} && \otimes && \text{I} && \otimes && \sigma^1 && \otimes && \sigma^1 &\quad ;\\
    &H_{13}&&= && \sigma^1 && \otimes && \text{I} && \otimes && \text{I} && \otimes && \text{I} && \otimes && \sigma^1 && \otimes && \text{I} && \otimes && \text{I} &\quad; \\
    &H_{14}&&= && \sigma^1 && \otimes && \text{I} && \otimes && \text{I} && \otimes && \text{I} && \otimes && \sigma^1 && \otimes && \text{I} && \otimes && \sigma^1 &\quad; \\
    &H_{15}&&= && \sigma^1 && \otimes && \text{I} && \otimes && \text{I} && \otimes && \text{I} && \otimes && \sigma^1 && \otimes && \sigma^1 && \otimes && \text{I} &\quad; \\
    &H_{16}&&= && \sigma^1 && \otimes && \text{I} && \otimes && \text{I} && \otimes && \text{I} && \otimes && \sigma^1 && \otimes && \sigma^1 && \otimes && \sigma^1  &\quad; 
\end{align*}

\section{Gallery}

\begin{figure}[ht!]
    \centering
    \includegraphics[width=0.44\linewidth]{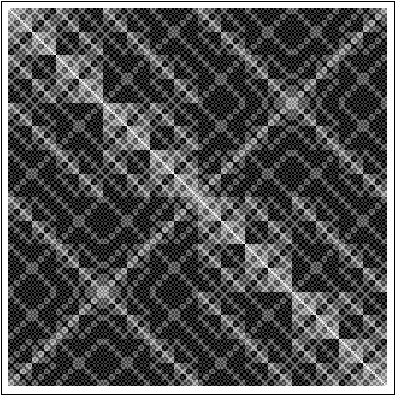}
    \includegraphics[width=0.44\linewidth]{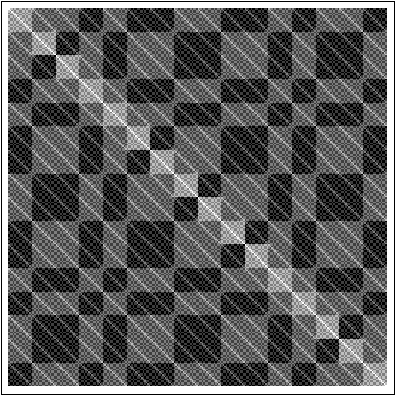}
    \includegraphics[width=0.44\linewidth]{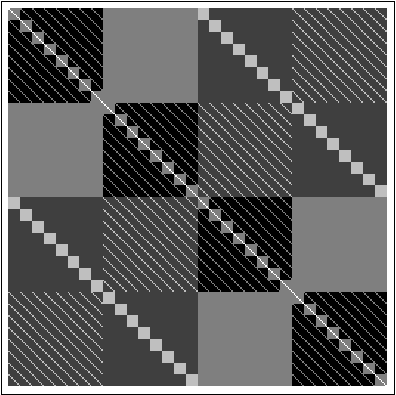}
    \caption{Distance Matrices of the topologies $E_8\times E_8$, $D_{16}$, and Spinning Particle Multiplet.}
    \label{fig:enter-label}
\end{figure}

\newpage

\bibliographystyle{hephys}
\bibliography{N16genomics}

\end{document}